\documentclass[12pt, final]{article}

\usepackage{virgil_env} 

\usepackage{fullpage}

\usepackage{nips10submit_e}
\nipsfinalcopy
\usepackage{nips10submit_e}
\nipsfinalcopy

\definecolor{mygray}{RGB}{153,153,153}
\definecolor{myred}{RGB}{221,77,57}
\definecolor{myblue}{RGB}{68,104,252}
\definecolor{mygreen}{RGB}{86,149,22}

\renewcommand*{\P}{\ensuremath{\mathbf{P}}\xspace}

\newcommand*{\bin}[1]{\texttt{#1}}

\newcommand*{\setX}{\ensuremath{\mathbf{X}}\xspace}

\newcommand*{\setS}{\ensuremath{\mathbf{S}}\xspace}
\newcommand*{\setP}{\ensuremath{\mathbf{P}}\xspace}

\newcommand*{\IbS}{\ensuremath{\textsf{IbE}}\xspace}
\newcommand*{\IbP}{\ensuremath{\textsf{IbDp}}\xspace}
\newcommand*{\IbPTwo}{\ensuremath{\textsf{IbB}}\xspace}
\newcommand*{\IbTwo}{\ensuremath{\textsf{Ib2p}}\xspace}
\newcommand*{\IbC}{\ensuremath{\textsf{Ib2p}}\xspace}
\newcommand*{\IbD}{\ensuremath{\textsf{IbAp}}\xspace}

\newcommand{\GP}{$\mathbf{(GP)}$\xspace}
\newcommand{\Eq}{$\mathbf{(Eq)}$\xspace}

\newcommand{\Mzero}{$\mathbf{\left(M_0\right)}$\xspace}
\newcommand{\Szero}{$\mathbf{\left(S_0\right)}$\xspace}

\newcommand{\SR}{$\mathbf{(SR)}$\xspace}

\usepackage{authblk}
\author{Virgil Griffith}
\author{Jonathan Harel}
\affil{Computation and Neural Systems, Caltech, Pasadena, CA 91125\vspace{.1in}}

\usepackage{wasysym}

\title{Irreducibility is Minimum Synergy Among Parts}

\newtheorem{lem}{Lemma}

\newcommand*{\Icupe}[2]{\ensuremath{\opname{I}_{\cup}\!\left( #1 \! : \! #2 \right)}}

\newcommand{\Xn}{\ensuremath{X_{1 \ldots n}}}

\renewcommand*{\vee}{\curlyvee}
\renewcommand*{\wedge}{\curlywedge}

\begin{document}

\maketitle
\vspace{-0.2in} \TODO{\today}

\begin{abstract}
For readers already familiar with Partial Information Decomposition (PID), we show that PID's definition of synergy enables quantifying at least four different notions of irreducibility. First, we show four common notions of ``parts'' give rise to a spectrum of four distinct measures of irreducibility. Second, we introduce a nonnegative expression based on PID for each notion of irreducibility. Third, we delineate these four notions of irreducibility with exemplary binary circuits.  This work will become more useful once the complexity community has converged on a palatable $\operatorname{I}_{\cap}$ or $\operatorname{I}_{\cup}$ measure.
\end{abstract}

\section{Introduction}
\label{sect:intro}
Irreducibility quantifies ``groupness'' or how much a group of agents acts as a ``single entity''.  By definition, a group of two or more agents irreducibly perform a task if and only if the performance of that task decreases when the agents work ``separately'', or in parallel isolation.  It's important to remember that it's the collective \emph{action} that is irreducible, not the agents themselves.  A concrete example of of irreducibility is the ``agents'' hydrogen and oxygen working to extinguish fire.  Even when $\opname{H}_2$ and $\opname{O}_2$ are both present in the same container, working separately neither extinguishes fire (on the contrary fire grows!).  But hydrogen and oxygen fused or ``grouped'' into a single entity, $\opname{H_{2}O}$, readily extinguishes fire.  In our work the agents are the $n$ predictors $X_1, \ldots, X_n$ and their collective action is predicting a single target r.v. $Y$.

Inspired by the $\phi$ measure\cite{balduzzi-tononi-08} which quantifies the minimum synergy beyond all partitions of disjoint parts, our work here shoes that the mathematics underlying the PID notion of synergy\cite{plw-10,qsmi,Iwedge} readily extends to quantifying \emph{irreducibility} simply by grouping together the elementary predictors into ``parts'', i.e., joint random variables.

One pertinent application of quantifying irreducibility is finding the most useful granularity for analyzing a complex system in which interactions occur at multiple scales.  Prior work \cite{bell03, bratko03, anas2007} has proposed measures of irreducibility, but they have various confounds\cite{qsmi}.

\section{Preliminaries}
\subsection{Informational Partial Order and Equivalence}

We assume an underlying probability space on which we define random variables denoted by capital letters (e.g., $X$, $Y$, and $Z$).  In this paper, we consider only random variables taking values on finite spaces.

Given random variables $X$ and $Y$, we write $X\preceq Y$ to signify that  there exists a measurable function $f$ such that $X=f(Y)$ almost surely (i.e., with probability one). In this case, following the terminology in \cite{li11}, we say that $X$ is \emph{informationally poorer} than $Y$; this induces a partial order on the set of random variables.  Similarly, we write $X \succeq Y$ if $Y \preceq X$, in which case we say $X$ is \emph{informationally richer} than $Y$.

If $X$ and $Y$ are such that $X\preceq Y$ and $X\succeq Y$, then we write $X\cong Y$.  In this case, again following \cite{li11}, we say that $X$ and $Y$ are \emph{informationally equivalent}.  In other words, $X \cong Y$ if and only if one can relabel the values of $X$ to obtain a random value that is equal to $Y$ almost surely, and vice versa.

This ``information-equivalence'' relation is an equivalence relation, so that we can partition the set of all random variables into disjoint equivalence classes. The $\preceq$ ordering is invariant within these equivalence classes in the following sense. If $X\preceq Y$ and $Y\cong Z$, then $X\preceq Z$. Similarly, if $X\preceq Y$ and $X\cong Z$, then $Z\preceq Y$. Moreover, within each equivalence class, the entropy is invariant.

\subsection{Information Lattice}

Next, we follow \cite{li11} and consider the \emph{join} and \emph{meet}
operators. These operators were defined for
\emph{information elements}, which are $\sigma$-algebras, or,
equivalently, equivalence classes of random variables. We deviate from
\cite{li11}, though, by defining the join and meet operators for random
variables, but we preserve their conceptual properties.

Given random variables $X$ and $Y$, we define $X\vee Y$ (called the
\emph{join} of $X$ and $Y$) to be an informationally poorest
(``smallest'' in the sense of the partial order $\preceq$) random variable  such that $X \preceq X \vee Y$ and $Y \preceq X \vee Y$.  In other words, if $Z$ is such that $X \preceq Z$ and $Y \preceq Z$, then $X \vee Y \preceq Z$.
Note that $X\vee Y$ is unique only up to equivalence with respect to $\cong$. In other words, $X\vee Y$ does not define a specific, unique random variable. Nonetheless, standard information-theoretic quantities are invariant over the set of random variables satisfying the condition specified above. For example, the entropy of $X\vee Y$ is invariant over the entire equivalence class of random variables satisfying the condition above. Similarly, the inequality $Z\preceq X\vee Y$ does not depend on the specific random variable chosen, as long as it satisfies the condition above. Note that the pair $(X,Y)$ is an instance of $X\vee Y$.

In a similar vein, given random variables $X$ and $Y$, we define $X\wedge Y$ (called the \emph{meet} of $X$ and $Y$) to be an informationally richest random variable (``largest'' in the sense of $\succeq$) such that $X\wedge Y\preceq X$ and $X\wedge Y\preceq Y$. In other words, if $Z$ is such that $Z\preceq X$ and $Z\preceq Y$, then $Z\preceq X\wedge Y$. Following \cite{gacs73}, we also call $X\wedge Y$ the \emph{common random variable} of $X$ and $Y$.  Again, considering the entropy of $X\wedge Y$ or the inequality $Z\preceq X\wedge Y$ does not depend on the specific random variable chosen, as long as it satisfies the condition above.

\subsection{Invariance and Monotonicity of Entropy}
Let $\ent{\cdot}$ represent the entropy function, and $\ent{\cdot|\cdot}$ the conditional entropy.  Chapter 3 established the invariance and monotonicity of the entropy and conditional entropy functions with respect to $\cong$ and $\preceq$.  From \cite{Iwedge}, the following hold:
\begin{itemize}
\item[(a)] If $X\cong Y$, then $\ent{X}=\ent{Y}$,
$\ent{X\middle|Z}=\ent{Y\middle|Z}$, and $\ent{Z\middle|X} = \ent{Z\middle|Y}$.
\item[(b)] If $X\preceq Y$, then $\ent{X}\leq\ent{Y}$,
$\ent{X\middle|Z}\leq\ent{Y\middle|Z}$, and $\ent{Z\middle|X}\geq\ent{Z\middle|Y}$.
\item[(c)] $X\preceq Y$ if and only if $\ent{X\middle|Y}=0$.
\end{itemize}

\subsection{Notation}
In our treatment of irreducibility, the $n$ agents are random variables $\{X_1, \ldots, X_n\}$, and the collective action the agents perform  is predicting (having mutual information about) a single target random variable $Y$.  We use the following notation throughout.  Let,

\begin{itemize}
    \item[$\setX$:] The set of $n$ elementary random variables (r.v.). $\setX \equiv \{X_1, X_2, \ldots, X_n\}$.  $n \geq 2$.
    \item[$\Xn$:] The \emph{whole}, the joint r.v. (cartesian product) of all $n$ elements, $\Xn \equiv X_1 \vee \cdots \vee X_n$.
    \item[$Y$:] The ``target'' random variable to be predicted.
    \item[$\mathcal{P}(\setX)$:] The set of all parts (random variables) derivable from a proper subset of $\setX$. From a set of $n$ elements there are $2^n - 2$ possible parts.  Formally, \newline $\mathcal{P}(\setX) \equiv \left\{ S_1 \vee \cdots \vee S_{|\setS|} : \setS \subset \setX, \setS \not= \emptyset \right\}$.
    \item[$\mathbf{P}$:] A set of $m$ parts $\mathbf{P}\equiv \{P_1, P_2, \ldots, P_m\}$, $2 \leq m \leq n$.  Each part $P_i$ is an element (random variable) in the set $\mathcal{P}(\setX)$.  The joint random variable of all $m$ parts is always informationally equivalent to $\Xn$, i.e., $P_1 \vee \cdots \vee P_m \cong \Xn$.  Hereafter, the terms ``part'' and ``component'' are used interchangeably.

    \item[$A_i$:] The $i$'th ``Almost''.  An ``Almost'' is a part (joint random variable) only lacking the element $X_i$. $1 \leq i \leq n$.  Formally, $A_i \equiv X_1 \vee \cdots \vee X_{i-1} \vee X_{i+1} \vee \cdots \vee X_n$.  
\end{itemize}

All capital letters are random variables.  All bolded capital betters are sets of random variables.



\section{Four common notions of irreducibility}
\label{sect:fourconceptions}

Prior literature \cite{chechik01,anas2007,bell03,berry03} has intuitively conceptualized the irreducibility of the information a whole $\Xn$ conveys about $Y$ in terms of how much information about $Y$ is lost upon ``breaking up'' $\Xn$ into a set of parts $\mathbf{P}$.  We express this intuition formally by computing the aggregate information $\mathbf{P}$ has about $Y$, and then subtracting it from the mutual information $\info{\Xn}{Y}$.  But what are the parts $\mathbf{P}$?  The four most common choices are:

\begin{enumerate}
	\item \textbf{The singleton elements}.  We take the set of $n$ elements, $\setX$, compute the mutual information with $Y$ when all $n$ elements work separately, and then subtract it from $\info{\Xn}{Y}$.  Information beyond the Elements ($\IbS$) is the weakest notion of irreducibility.  In the PI-diagram\cite{qsmi} of $\info{\Xn}{Y}$, $\IbS$ is the sum of all synergistic PI-regions.

	\item \textbf{Any partition of (disjoint) parts}.  We enumerate all possible partitions of set $\setX$.  Formally, a partition $\mathbf{P}$ is any set of parts $\{P_1, \ldots, P_m\}$ such that, $P_i \wedge P_j \prec X_k$ where $i,j \in \{1, \ldots, m\}$, $i \not= j$, and $k \in \{1, \ldots, n\}$.  For each partition, we compute the mutual information with $Y$ when its $m$ parts work separately.  We then take the maximum information over all partitions and subtract it from $\info{\Xn}{Y}$.  Information beyond the Disjoint Parts ($\IbP$) quantifies $\info{\Xn}{Y}$'s irreducibility to information conveyed by disjoint parts.
	
	\item \textbf{Any two parts}. We enumerate all ``part-pairs'' of set $\setX$.  Formally, a part-pair $\mathbf{P}$ is any set of exactly two elements in $\mathcal{P}(\setX)$.  For each part-pair, we compute the mutual information with $Y$ when the parts work separately.  We then take the maximum mutual information over all part-pairs and subtract it from $\info{\Xn}{Y}$.  Information beyond the Two Parts ($\IbTwo$) quantifies $\info{\Xn}{Y}$'s irreducibility to information conveyed by any pair of parts.
		
	\item \textbf{All possible parts}.  We take the set of all possible parts of set $\setX$, $\mathcal{P}(\setX)$, and compute the information about $Y$ conveyed when all $2^{n} - 2$ parts work separately and subtract it from $\info{\Xn}{Y}$.  Information beyond All Parts ($\IbD$) is the strongest notion of irreducibility.  In the PI-diagram of $\info{\Xn}{Y}$, $\IbD$ is the value of PI-region $\{1\ldots n\}$.  
\end{enumerate}

\section{Quantifying the four notions of irreducibility}
\label{sect:defining}
To calculate the information in the whole beyond its elements, the first thing that comes to mind is to take the whole and subtract the sum over the elements, i.e., $\info{\Xn}{Y} - \sum_{i=1}^{n} \info{X_i}{Y}$.  However, the sum \emph{double-counts} when over multiple elements convey the same information about $Y$.  To avoid double-counting the same information, we need to change the sum to ``union''.  Whereas summing adds duplicate information multiple times, unioning adds duplicate information only once.  This guiding intuition of ``whole minus union'' leads to the definition of irreducibility as the information conveyed by the whole minus the ``union information'' over its parts.

We provide expressions for $\IbS$, $\IbP$, $\IbC$, and $\IbD$ for arbitrary $n$.  All four equations are the information conveyed by the whole, $\info{\Xn}{Y}$, minus the maximum union information about $Y$ over some parts $\P$, $\Icupe{P_1, \ldots, P_m}{Y}$.  There are currently several candidate definitions of the union information\cite{qsmi,Iwedge,polani12,bertschinger12}, but for our four irreducibility measures to work all that is required is that the $\Icup$ measure satisfy:

\begin{itemize}
    \item[\GP] Global Positivity: $\Icupe{P_1,\ldots,P_m}{Y} \geq 0$,
and $\Icupe{P_1,\ldots,P_m}{Y} = 0$ if $Y$ is a constant.

\item[\Eq] Equivalence-Class Invariance:
$\Icupe{P_1,\ldots,P_m}{Y}$ is invariant under substitution of $P_i$ (for any $i=1,\ldots,m$) or $Y$ by an informationally equivalent random variable.

    \item[\Mzero] Weak Monotonicity:  $\Icupe{P_1,\ldots,P_m, W}{Y} \geq \Icupe{P_1,\ldots,P_m}{Y}$ with equality if there exists $P_i \in \{P_1, \ldots, P_m\}$  such that $W \preceq P_i$.

    \item[\Szero] Weak Symmetry:
$\Icupe{P_1,\ldots,P_m}{Y}$ is invariant under reordering of $P_1, \ldots, P_m$.

    \item[\SR] Self-Redundancy: $\Icupe{P_1}{Y} = \info{P_1}{Y}$. The union information a single part $P_1$ conveys about the target $Y$ is equal to the mutual information between $P_1$ and the target.

    \item[$\mathbf{(UB)}$] Upperbound:  $\Icupe{P_1, \ldots, P_m}{Y} \leq \info{P_1 \vee \cdots \vee P_m}{Y}$.  In this particular case, the joint r.v. $P_1 \vee \cdots \vee P_m \cong \Xn$, so this equates to $\Icupe{P_1, \ldots, P_m}{Y} \leq \info{\Xn}{Y}$.
\end{itemize}

\subsection{Information beyond the Elements}
\label{sect:definingseparation}

Information beyond the Elements, $\IbS( \setX : Y )$ quantifies how much information in $\info{\Xn}{Y}$ isn't conveyed by any element $X_i$ for $i \in \{1, \ldots, n\}$.  The Information beyond the Elements is,
\begin{equation}
\label{eq:synergydef}	\IbS( \setX : Y ) \equiv \info{\Xn}{Y} - \opI_{\cup}\!\left( X_1, \ldots, X_n : Y \right) \; .
\end{equation}

Information beyond the Elements, or \emph{synergistic mutual information}\cite{qsmi}, quantifies the amount of information in $\info{\Xn}{Y}$ that \emph{only coalitions} of elements convey.

\subsection{Information beyond Disjoint Parts: $\IbP$}
\label{sect:definingpartition}

Information beyond Disjoint Parts, $\IbP( \setX : Y )$, quantifies how much information in $\info{\Xn}{Y}$ isn't conveyed by any partition of set $\setX$.  Like $\IbS$, $\IbP$ is the total information minus the ``union information'' over components.  Unlike $\IbS$, the components are not the $n$ elements but the parts of a partition. Some algebra proves that the partition with the maximum mutual information will always be a bipartition; thus we can safely restrict the maximization to bipartitions.\footnote{See Appendix A for a proof.}  Therefore to quantify $\info{\Xn}{Y}$'s irreducibility to disjoint parts, we maximize over all $2^{n-1} - 1$ bipartitions of set $\setX$.  Altogether, the Information beyond Disjoint Parts is,
\begin{eqnarray}
	\IbP\!\left( \setX : Y \right) &\equiv& \info{\Xn}{Y} - \max_{\substack{P_1 \in \mathcal{P}(\setX) \vspace{-5pt} \\ \vdots \\ P_m \in \mathcal{P}(\setX) \\ P_i \wedge P_j \prec X_k, \ \forall i\not= j \ k \in \{1, \ldots, n\} }} \Icupe{P_1,\ldots, P_m}{Y} \\
	&=& \info{\Xn}{Y} - \max_{S \in \mathcal{P}(\setX)} \  \opI_{\cup}\!\left( S, \setX \setminus S  : Y \right) \; .
\end{eqnarray}

\subsection{Information beyond Two Parts: $\IbC$}
\label{sect:subsystems}

Information beyond Two Parts, $\IbC( \setX : Y )$, quantifies how much information in $\info{\Xn}{Y}$ isn't conveyed by any pair of parts.  Like $\IbP$, $\IbC$ subtracts the maximum union information over two parts.  Unlike $\IbP$, the two parts aren't disjoint.  Some algebra proves that the part-pair conveying the most information about $Y$ will always be a pair of ``Almosts''.\footnote{See Appendix A for a proof.}  Thus we can safely restrict the maximization over all pairs of Almosts, and we maximize over the ${n \choose 2} = \frac{n(n-1)}{2}$ pairs of Almosts. Altogether, the Information beyond Two Parts is,
\begin{eqnarray}
\label{eq:Ib2def}
\IbC\!\left( X_1, \ldots, X_n : Y \right) &\equiv& \info{\Xn}{Y} - \max_{\substack{P_1 \in \mathcal{P}(\setX) \\ P_2 \in \mathcal{P}(\setX) }} \Icupe{P_1, P_2}{Y} \\
&=& \info{\Xn}{Y} - \max_{ \substack{ i, j \in \{1, \ldots, n\} \\ i \not= j}} \Icupe{A_i, A_j}{Y} \; .
\end{eqnarray}

\subsection{Information beyond All Parts: $\IbD$}
\label{sect:holism}

Information beyond All Parts, $\IbD( \setX : Y )$, quantifies how much information in $\info{\Xn}{Y}$ isn't conveyed by any part. Like $\IbC$, $\IbD$ subtracts the union information over overlapping parts.  Unlike $\IbC$, the union is not over two parts, but all possible parts.  Some algebra proves that the entirety of the information conveyed by all $2^n - 2$ parts working separately is equally conveyed by the $n$ Almosts working separately.\footnote{See Appendix A for a proof.}  Thus we can safely contract the union information to the $n$ Almosts.  Altogether, the Information beyond All Parts is,

\begin{equation}
\begin{split}
\label{eq:holismdef} \IbD \left( X_1, \ldots, X_n : Y \right) &\equiv \info{\Xn}{Y} - \Icupe{\mathcal{P}(\setX)}{Y} \\
    &= \info{\Xn}{Y} - \Icupe{A_1, A_2, \ldots, A_n}{Y} \; .
\end{split}    
\end{equation}

 Whereas Information beyond the Elements quantifies the amount of information in $\info{\Xn}{Y}$ only conveyed by coalitions, Information beyond All Parts, or \emph{holistic mutual information}, quantifies the amount of information in $\info{\Xn}{Y}$ only conveyed by the whole.

By properties (\textbf{GP}) and (\textbf{UB}), our four measures are nonnegative and bounded by $\info{\Xn}{Y}$.  Finally, each succeeding of notion of components is a generalization of the prior.  This successive generality gives rise to the handy inequality:
\begin{equation}
    \label{eq:ordering}
    \IbD( \setX : Y )	 \leq \IbC( \setX : Y ) \leq \IbP( \setX : Y ) \leq \IbS( \setX : Y )  \; .
\end{equation}


\section{Exemplary Binary Circuits}
\label{sect:exemplarycircuits}
\begin{figure}[h!bt]
	\centering
	\begin{minipage}[c]{0.26\linewidth} \centering \subfloat[$\Prob{x_1,x_2,y}$]{ \begin{tabular}{ c | c c} \cmidrule(r){1-2}
	$X_1$ $X_2$ &$Y$ \\
	\cmidrule(r){1-2} 
	\bin{0 0} & \bin{0} & \quad \nicefrac{1}{4}\\
	\bin{0 1} & \bin{1} & \quad \nicefrac{1}{4}\\
	\bin{1 0} & \bin{1} & \quad \nicefrac{1}{4}\\
	\bin{1 1} &  \bin{0} & \quad \nicefrac{1}{4}\\
	\cmidrule(r){1-2} 
	\end{tabular} }
	\end{minipage}
	\begin{minipage}[c]{0.365\linewidth} \centering	
	\subfloat[circuit diagram]{ \includegraphics[height=1.2in]{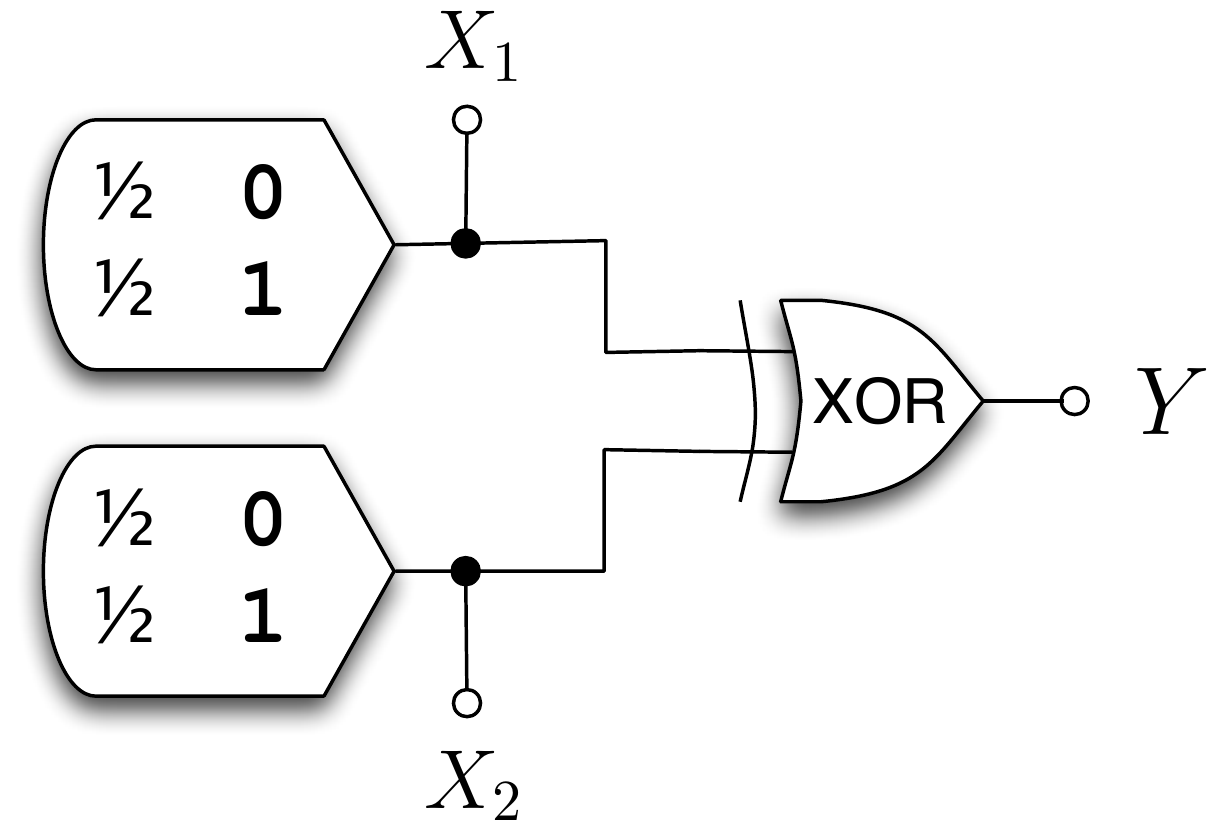} }
	\end{minipage}		
	\begin{minipage}[c]{0.36\linewidth} \centering
		\subfloat[PI-diagram]{ \includegraphics[height=1.0in]{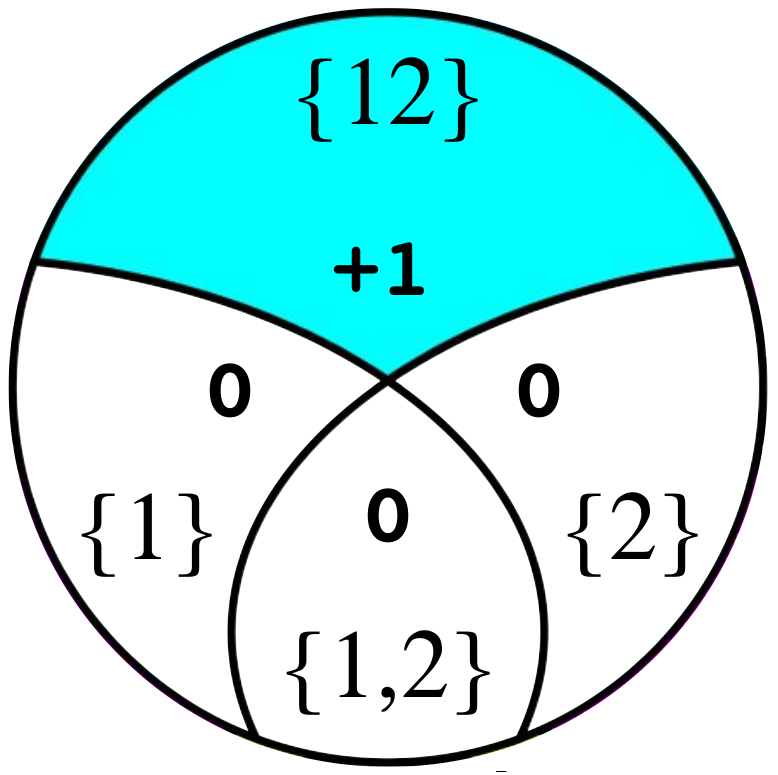} }
	\end{minipage}
	\caption{Example \textsc{Xor}.  $X_1 X_2$ irreducibly specifies $Y$.  $\info{X_1 X_2}{Y}~=~\ent{Y}~=~1$ bit.}
	\label{fig:XOR}
\end{figure}

\begin{figure}[h!bt]
\centering
	\subfloat[$\IbS(X_1, X_2 \: Y )$]{ \includegraphics[height=1.2in]{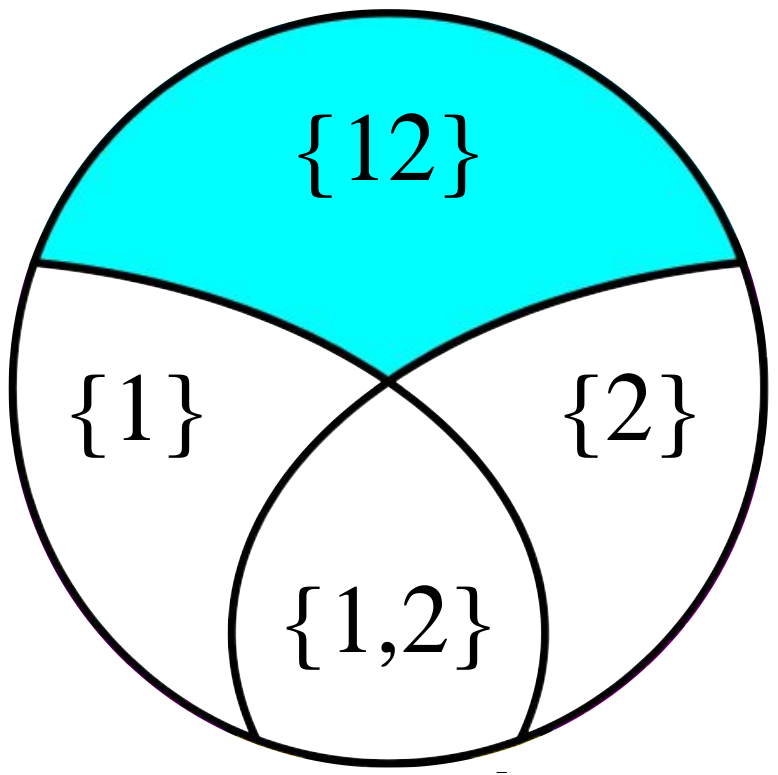} \label{fig:PID2_a} }
	\subfloat[$\IbP(X_1, X_2 \: Y )$]{ \includegraphics[height=1.2in]{PID2-syninfo-small.pdf} }
	\subfloat[$\IbC(X_1, X_2 \: Y )$]{ \includegraphics[height=1.2in]{PID2-syninfo-small.pdf} }
	\subfloat[$\IbD(X_1, X_2 \: Y )$]{ \includegraphics[height=1.2in]{PID2-syninfo-small.pdf} \label{fig:PID2_d} }

	\subfloat[$\IbS( X_1, X_2, X_3 : Y)$]{ \includegraphics[height=2.3in]{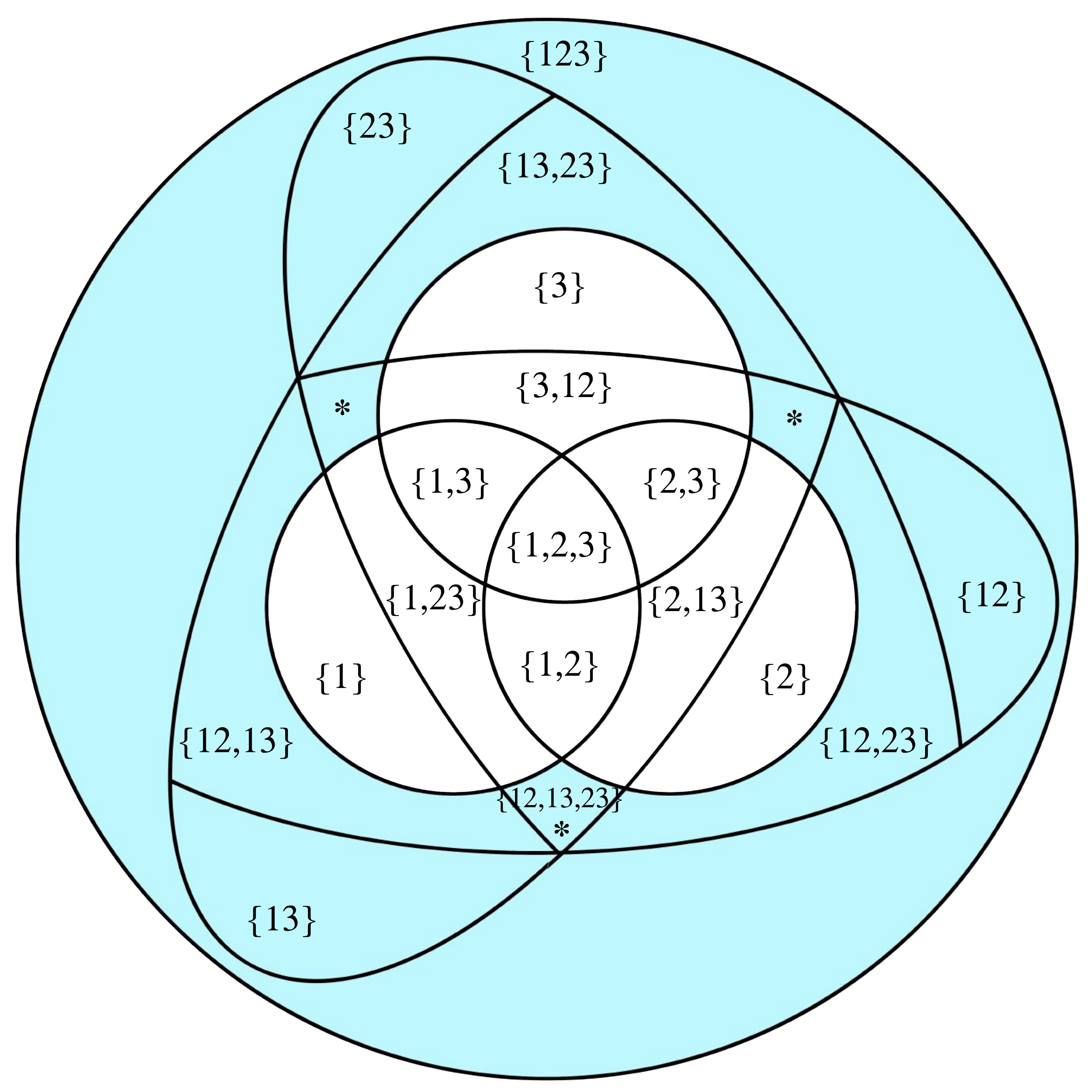} 
\label{fig:PID3_IbS} }
	\subfloat[$\IbD( X_1, X_2, X_3 : Y)$]{ \includegraphics[height=2.3in]{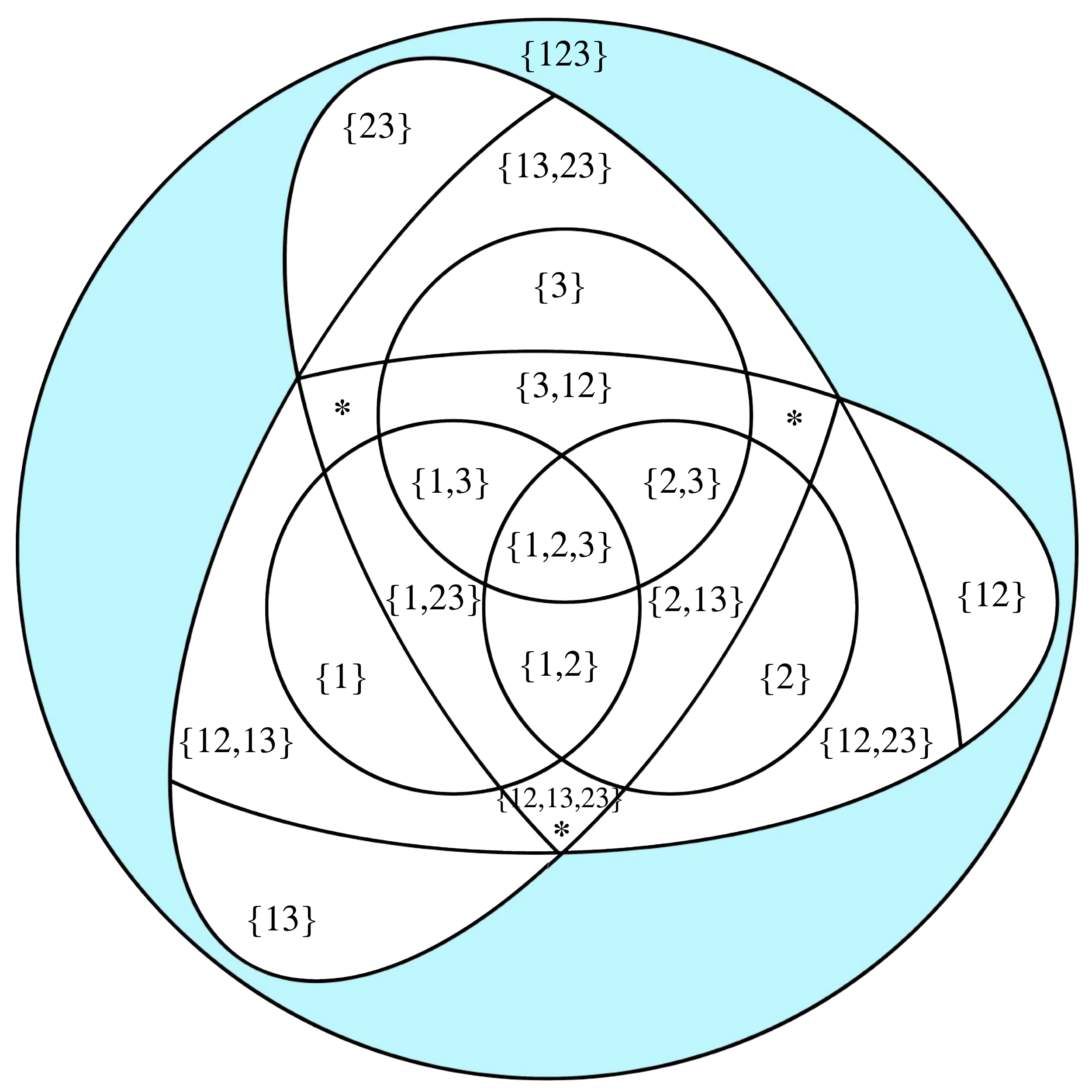} 
\label{fig:PID3_IbD} }

	\subfloat[$\mathbf{P} = \{X_1 X_2, X_3\}$]{ \includegraphics[height=1.9in]{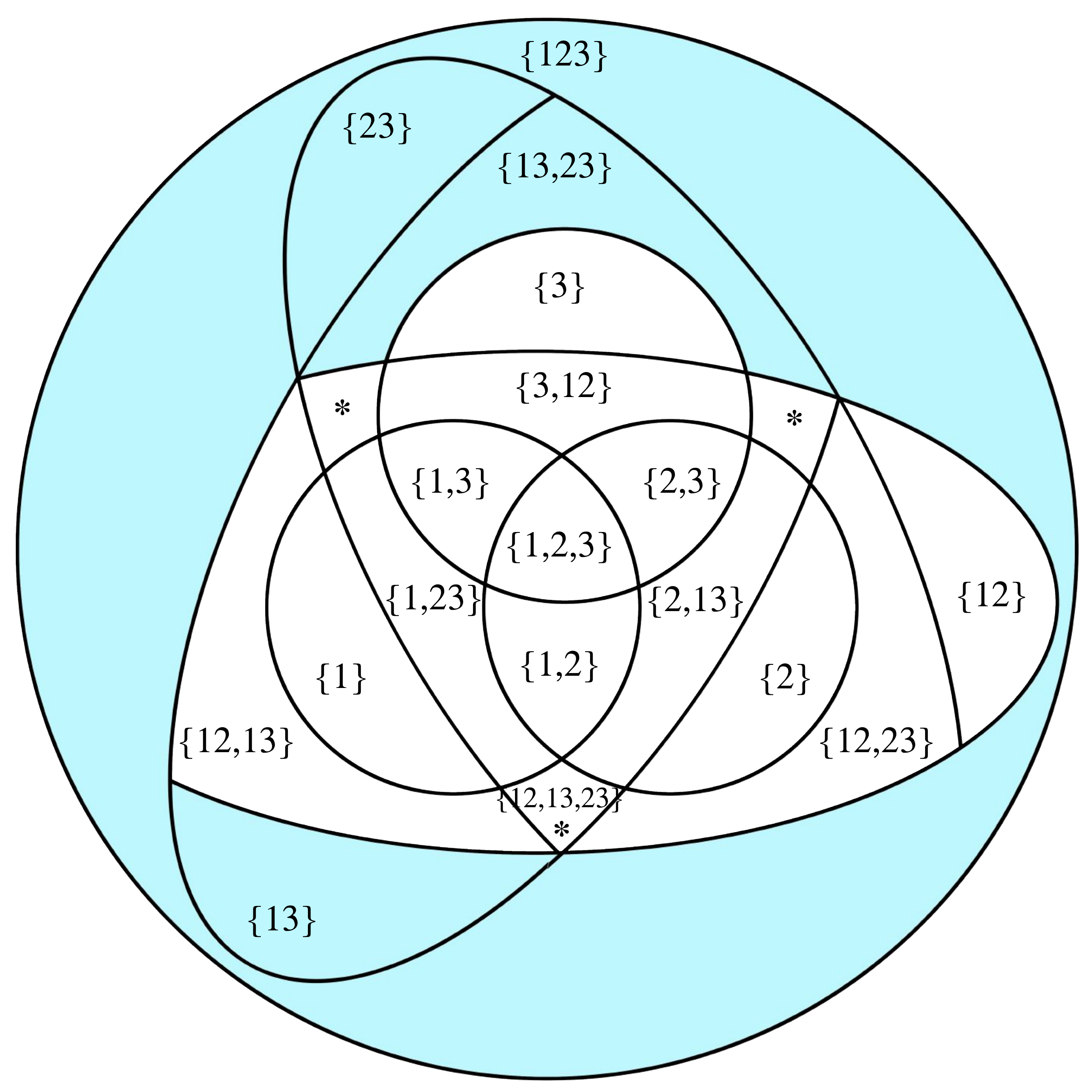} 
\label{fig:PID3_COPa} }	
	\subfloat[$\mathbf{P} = \{X_1 X_3, X_2\}$]{ \includegraphics[height=1.9in]{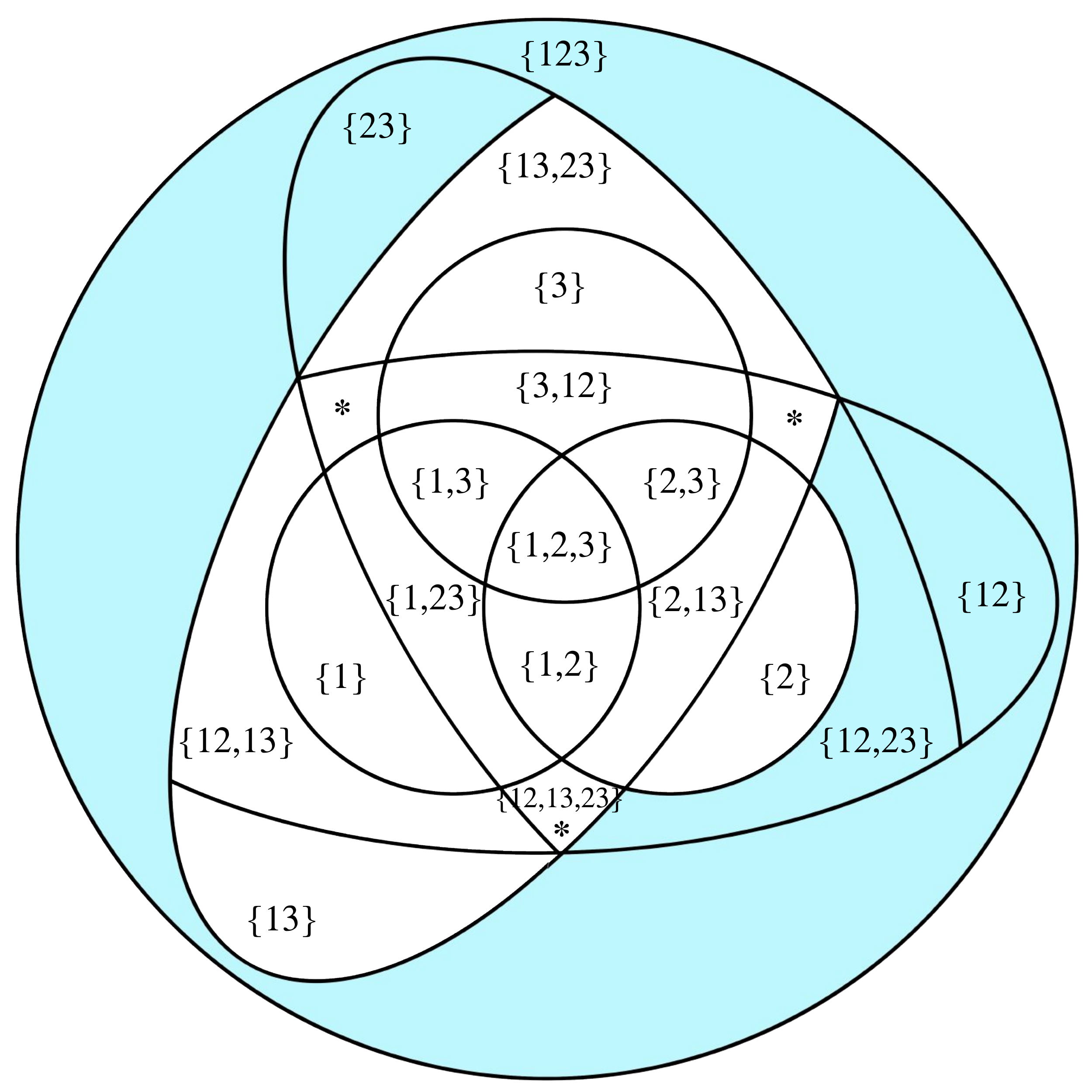} 
\label{fig:PID3_COPb} }
	\subfloat[$\mathbf{P} = \{X_2 X_3, X_1\}$]{ \includegraphics[height=1.9in]{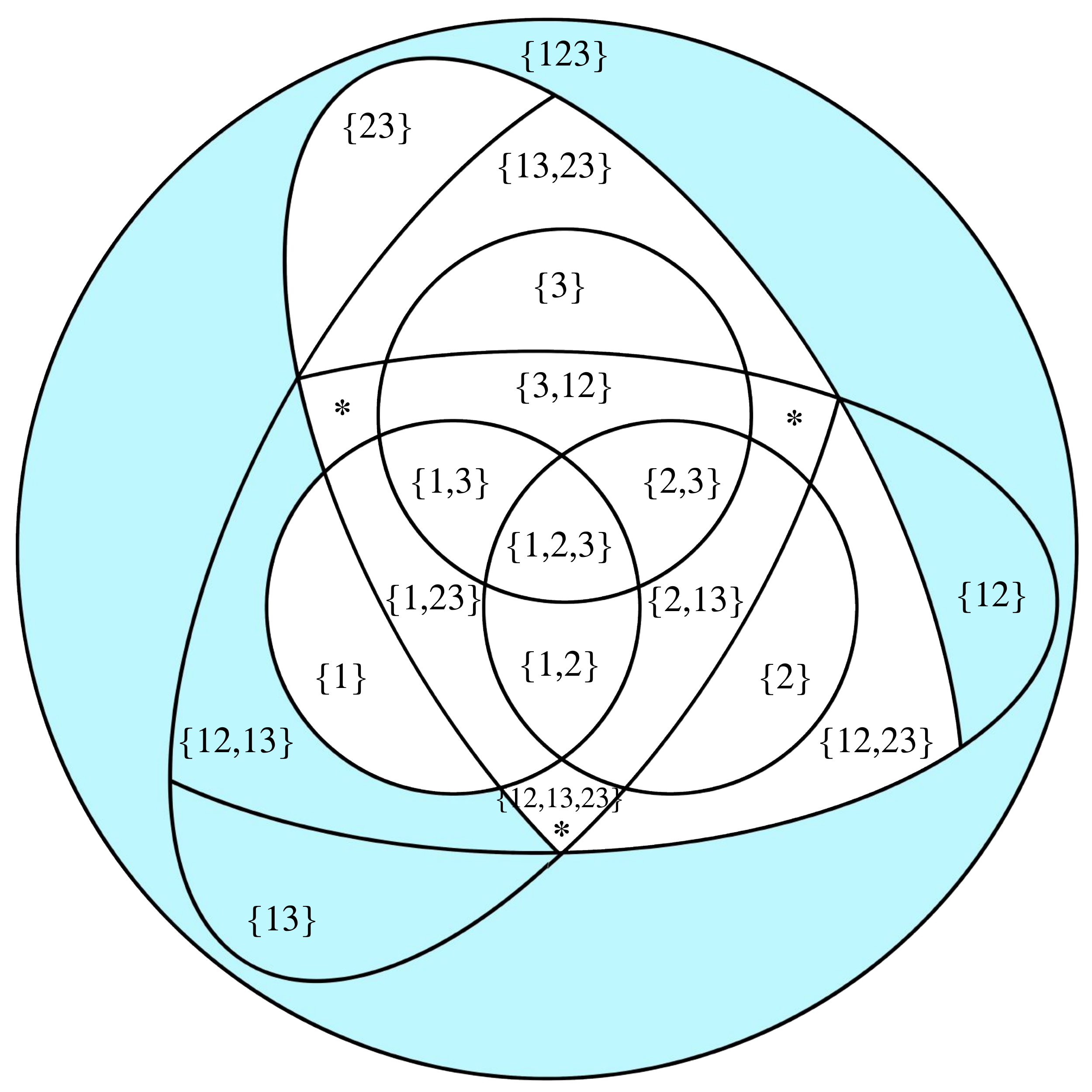} 
\label{fig:PID3_COPc} }	

	\subfloat[$\mathbf{P} = \{X_1 X_2, X_2 X_3\}$]{ \includegraphics[height=1.9in]{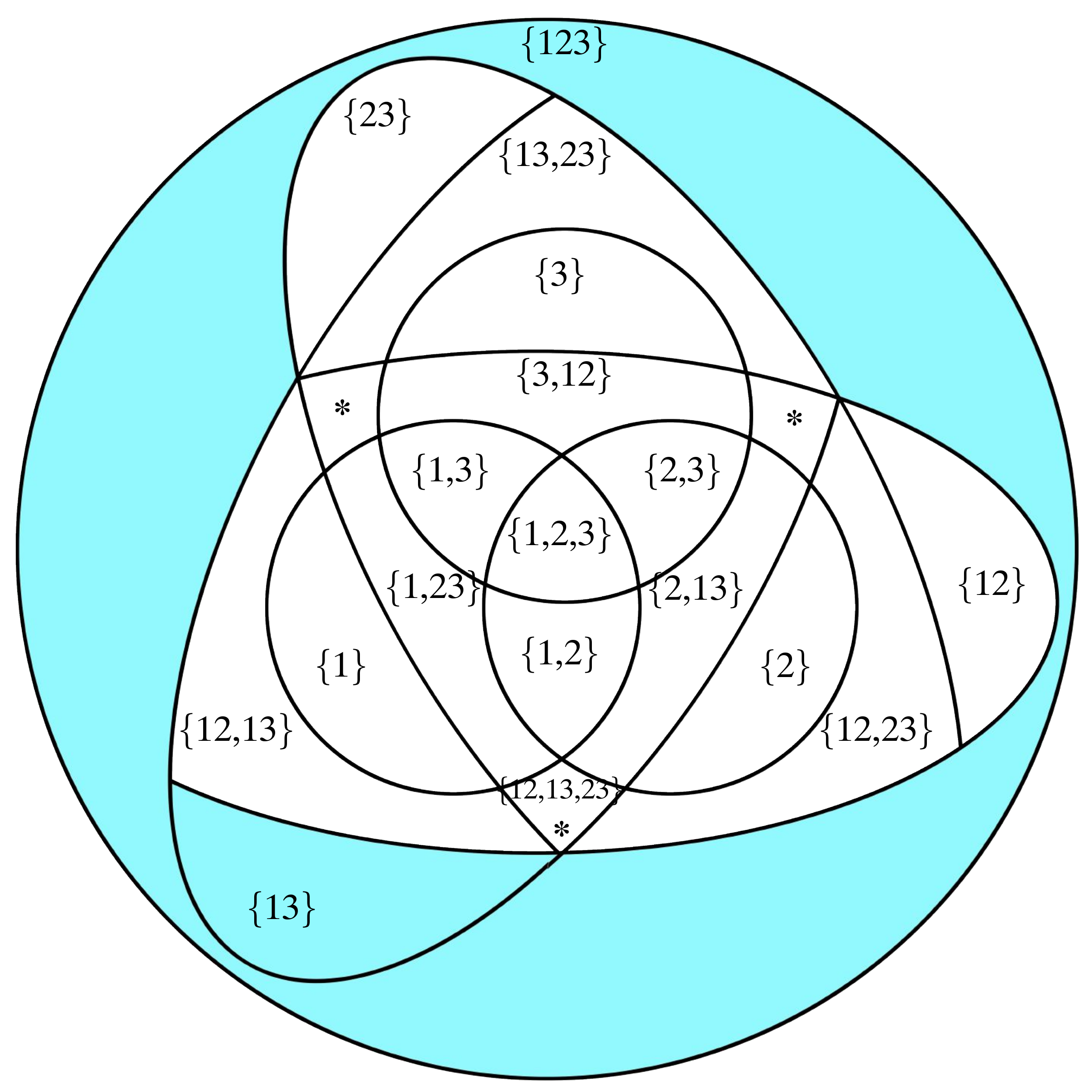} 
\label{fig:PID3_Ib2c} }	
	\subfloat[$\mathbf{P} = \{X_1 X_2, X_1 X_3\}$]{ \includegraphics[height=1.9in]{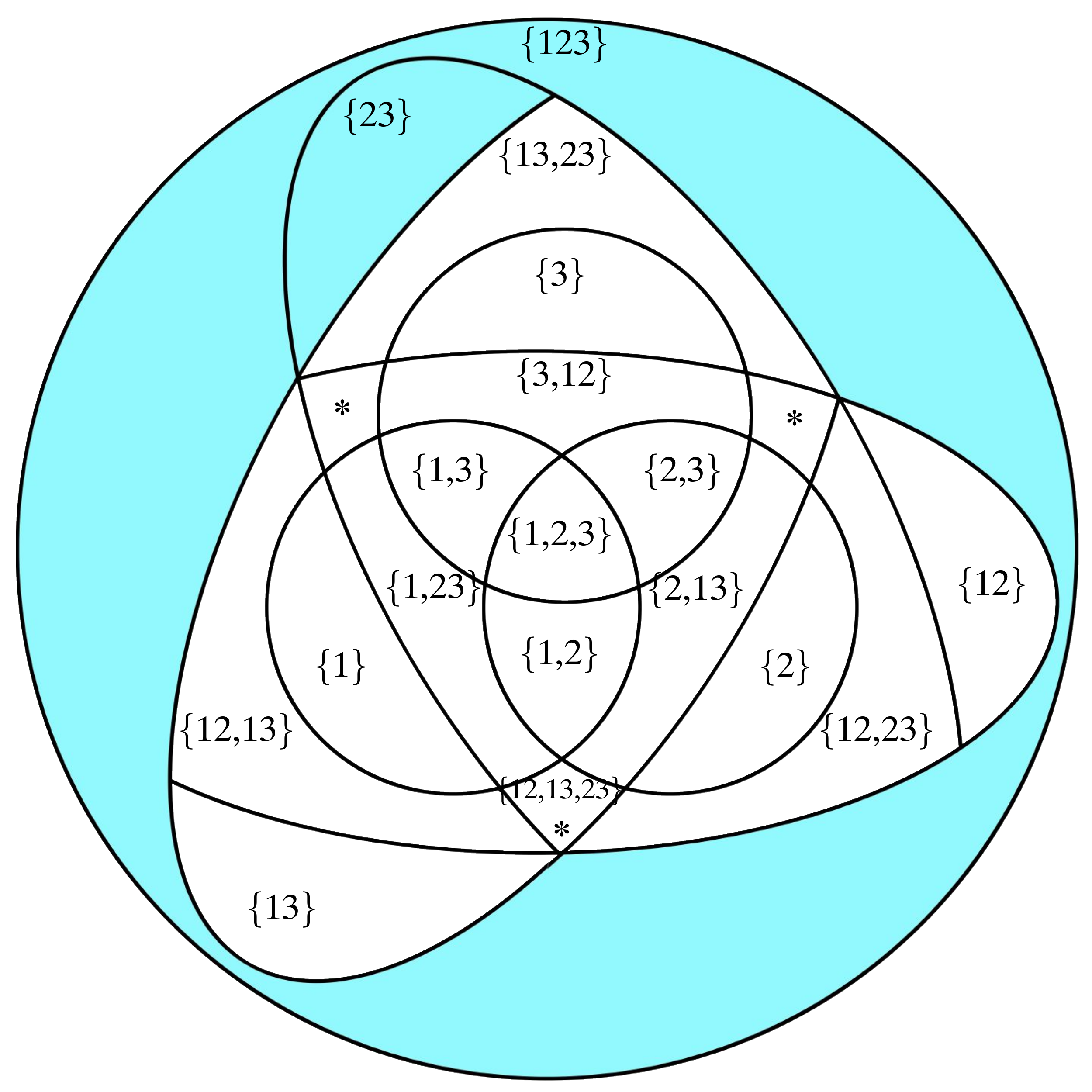} 
\label{fig:PID3_Ib2a} }	
	\subfloat[$\mathbf{P} = \{X_1 X_3, X_2 X_3\}$]{ \includegraphics[height=1.9in]{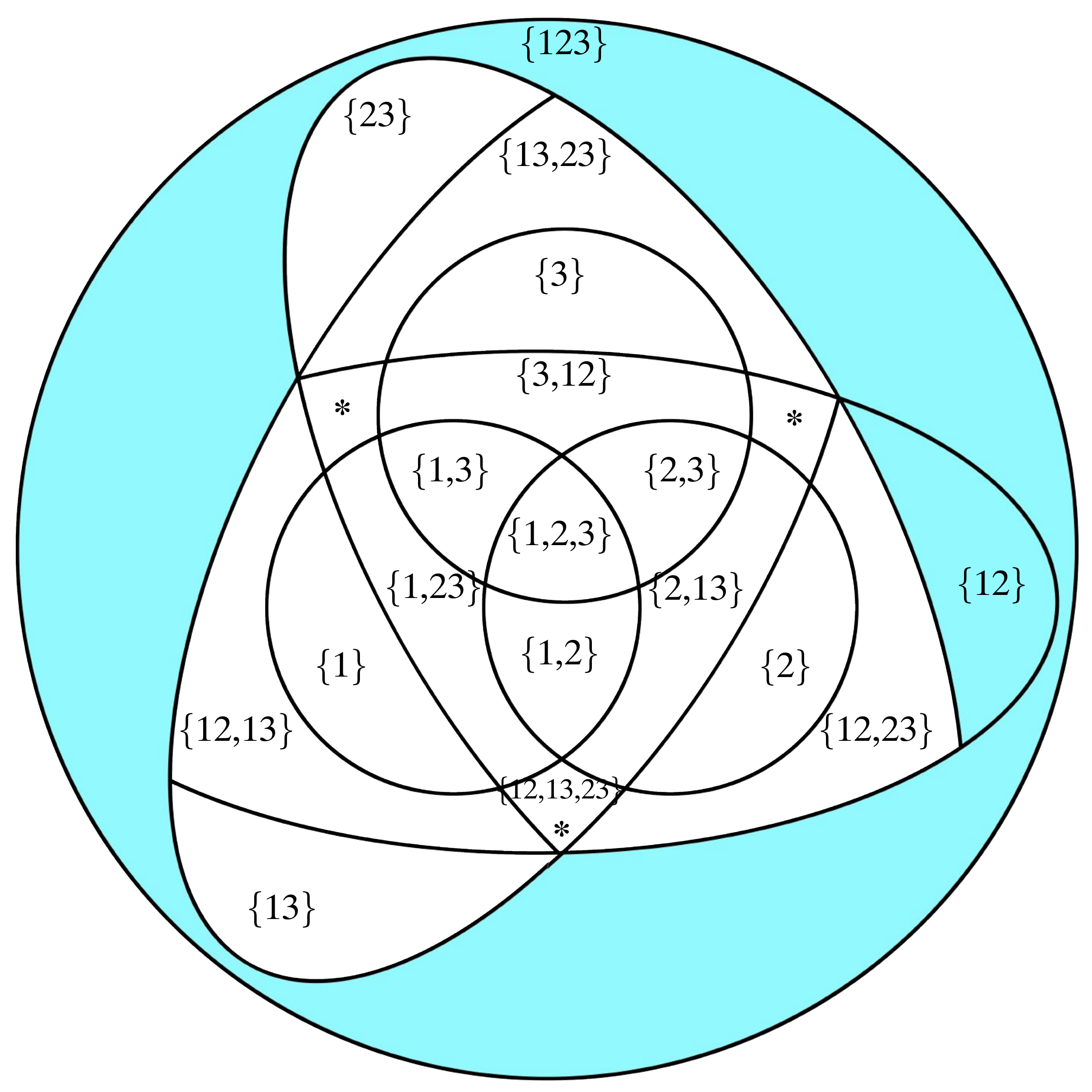} 
\label{fig:PID3_Ib2b} }

\caption{PI-diagrams depicting our four irreducibility measures when $n=2$ and $n=3$ in subfigures (a)--(d) and (e)--(l) respectively.  For $n=3$: $\IbS$ is (e), $\IbD$ is (f), $\IbP$ is the \emph{minimum} value over subfigures (g)--(i), and $\IbTwo$ is the \emph{minimum} value over subfigures (j)--(l).}
	\label{fig:CoP}
\end{figure}

For $n=2$, all four notions of irreducibility are equivalent---each one is simply the value of PI-region $\{12\}$ (see subfigures \ref{fig:PID2_a}--\subref*{fig:PID2_d}).  The canonical example of irreducibility for $n=2$ is example \textsc{Xor} (\figref{fig:XOR}).  In \textsc{Xor}, the irreducibility of $X_1$ and $X_2$ specifying $Y$ is analogous to irreducibility of hydrogen and oxygen extinguishing fire.  The whole $X_1 X_2$ fully specifies $Y$, $\info{X_1 X_2}{Y} = \ent{Y} = 1$ bit, but $X_1$ and $X_2$ separately convey nothing about $Y$, $\info{X_1}{Y} = \info{X_2}{Y} = 0$ bits.

For $n>2$, the four notions of irreducibility diverge; subfigures \ref{fig:PID3_IbS}--\subref*{fig:PID3_Ib2c} depicts $\IbS$, $\IbD$, $\IbP$, and $\IbTwo$ when $n=3$.  We provide exemplary binary circuits delineating each measure.  Every circuit has $n=3$ elements, meaning $\setX = \{X_1, X_2, X_3\}$, and build atop example \textsc{Xor}.

\subsection{XorUnique: Irreducible to elements, yet reducible to a partition}

To concretize how a collective action could be irreducible to elements yet still reducible to a partition, consider a hypothetical set of agents $\{ X_1, X_2, \ldots, X_{100} \}$ where the first 99 agents cooperate to specify $Y$, but agent $X_{100}$ doesn't cooperate with the joint random variable $X_{1} \cdots X_{99}$.  The $\IbS$ among these 100 agents would be \emph{positive}, however, $\IbP$ would be \emph{zero} because the work that $X_1 \cdots X_{100}$ performs can be reduced to two disjoint parts, $X_{1} \cdots X_{99}$ and $X_{100}$, working separately.

\begin{figure}[h!bt]
	\centering
	\begin{minipage}[c]{0.55\linewidth} \centering
	\subfloat[circuit diagram]{ \includegraphics[width=2.4in]{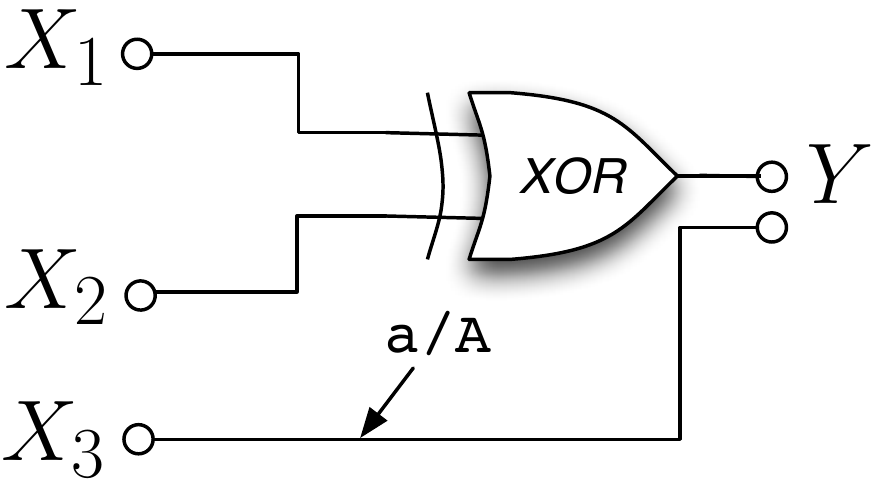} }
	\end{minipage}
	
	\begin{minipage}[c]{0.4\linewidth} \centering
	\subfloat[$\Prob{x_1, x_2, x_3, y}$]{ \begin{tabular}{ c | c c } \cmidrule(r){1-2}
$\ \, X_1 \, X_2 \, X_3$  &$Y$ \\
\cmidrule(r){1-2} 
\bin{0 0 a} & \bin{0a} & \quad \nicefrac{1}{8}\\
\bin{0 1 a} & \bin{1a} & \quad \nicefrac{1}{8}\\
\bin{1 0 a} & \bin{1a} & \quad \nicefrac{1}{8}\\
\bin{1 1 a} & \bin{0a} & \quad \nicefrac{1}{8}\\
\addlinespace
\bin{0 0 A} & \bin{0A} & \quad \nicefrac{1}{8}\\
\bin{0 1 A} & \bin{1A} & \quad \nicefrac{1}{8}\\
\bin{1 0 A} & \bin{1A} & \quad \nicefrac{1}{8}\\
\bin{1 1 A} & \bin{0A} & \quad \nicefrac{1}{8}\\
\cmidrule(r){1-2} 
\end{tabular} }
\end{minipage} \begin{minipage}[c]{0.55\linewidth} \centering
	\subfloat[PI-diagram]{ \includegraphics[height=2.3in]{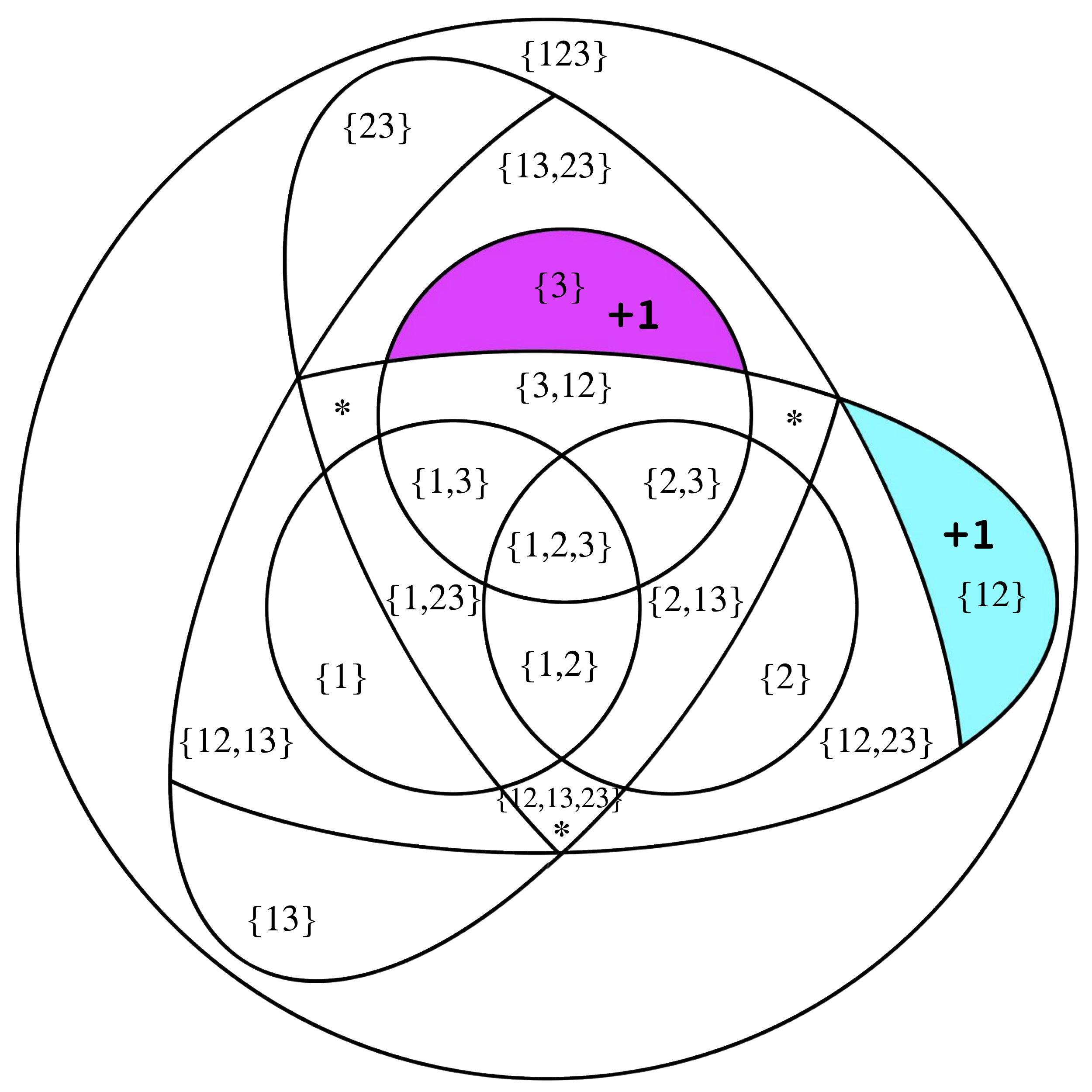} }
	\end{minipage}
	\caption{ Example \textsc{XorUnique}.  Target $Y$ has two bits of uncertainty.  The doublet $X_1 X_2$ specifies the ``digit bit'', and the singleton $X_3$ specifies the ``letter bit'' for a total of $\info{X_1 X_2}{Y}~+~\info{X_3}{Y}~=~\ent{Y}~=~2$ bits. $X_1 X_2 X_3$'s specification of $Y$ is irreducible to singletons yet fully reduces to the disjoint parts $\{X_1 X_2, X_3\}$.}
	\label{fig:XorUnique}
\end{figure}

Example \textsc{XorUnique} (\figref{fig:XorUnique}) is analogous to the situation above.  The whole specifies two bits of uncertainty, $\info{X_1 X_2 X_3}{Y}=\ent{Y}=2$ bits.  The doublet $X_1 X_2$ solely specifies the ``digit-bit'' of $Y$ (\bin{0}/\bin{1}), $\info{X_1 X_2}{Y}=1$ bit, and the singleton $X_3$ solely specifies the ``letter-bit'' of $Y$ (\bin{a}/\bin{A}), $\info{X_3}{Y}=1$ bit.  We apply each notion of irreducibility to \textsc{XorUnique}:

\begin{enumerate}
	\item[\textbf{\textsf{IbE}}] How much of $X_1 X_2 X_3$'s information about $Y$ can be reduced to the information conveyed by the singleton elements working separately?  Working alone, $X_3$ still specifies the letter-bit of $Y$, but $X_1$ nor $X_2$ can unilaterally specify the digit-bit of $Y$, $\info{X_1}{Y}~=~0$ and $\info{X_2}{Y}~=~0$ bits.  As only the letter-bit is specified when the three singletons work separately, $\IbS\left( \setX : Y \right) = \info{X_1 X_2 X_3}{Y} - 1 = 2 - 1 = 1$ bit.
	
	\item[\textbf{\textsf{IbDp}}] How much of $X_1 X_2 X_3$'s information about $Y$ can be reduced to the information conveyed by disjoint parts working separately?  Per subfigures \ref{fig:PID3_COPa}--\subref*{fig:PID3_COPc}, there are three bipartitions of $X_1 X_2 X_3$, and one of them is $\{X_1 X_2, X_3\}$.  The doublet part $X_1 X_2$ specifies the digit-bit of $Y$, and the singleton part $X_3$ specifies the letter-bit of $Y$.  As there is a partition of $X_1 X_2 X_3$ that fully accounts for $X_1 X_2 X_3$'s specification of $Y$, $\IbP\!\left( \setX : Y \right) = 2 - 2 = 0$ bits.
	
	\item[\textbf{\textsf{Ib2p/IbAp}}] How much of $X_1 X_2 X_3$'s information about $Y$ can be reduced to the information conveyed by two parts working separately?  From above we see that $\IbP$ is zero bits.  Per eq.~\eqref{eq:ordering}, $\IbTwo$ and $\IbD$ are stricter notions of irreducibility than $\IbP$, therefore $\IbTwo$ and $\IbD$ must also be zero bits.
\end{enumerate}

\subsection{DoubleXor: Irreducible to a partition, yet reducible to a pair}
\begin{figure}[tbh]
	\centering
\begin{minipage}[l]{0.4\linewidth}
	\vspace{0.73in}	
	\subfloat[circuit diagram]{ \includegraphics[width=2.2in]{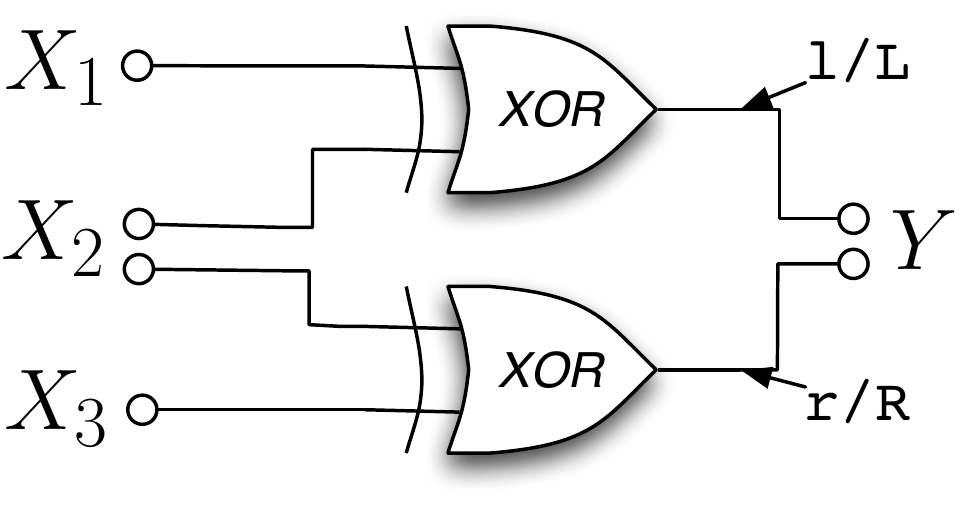} } \\
	\subfloat[$\Prob{x_1,x_2,x_3,y}$]{See Appendix B for the joint distribution. } \\
\end{minipage}	\begin{minipage}[r]{0.5\linewidth} \centering
	\subfloat[PI-diagram]{ \includegraphics[height=2.3in]{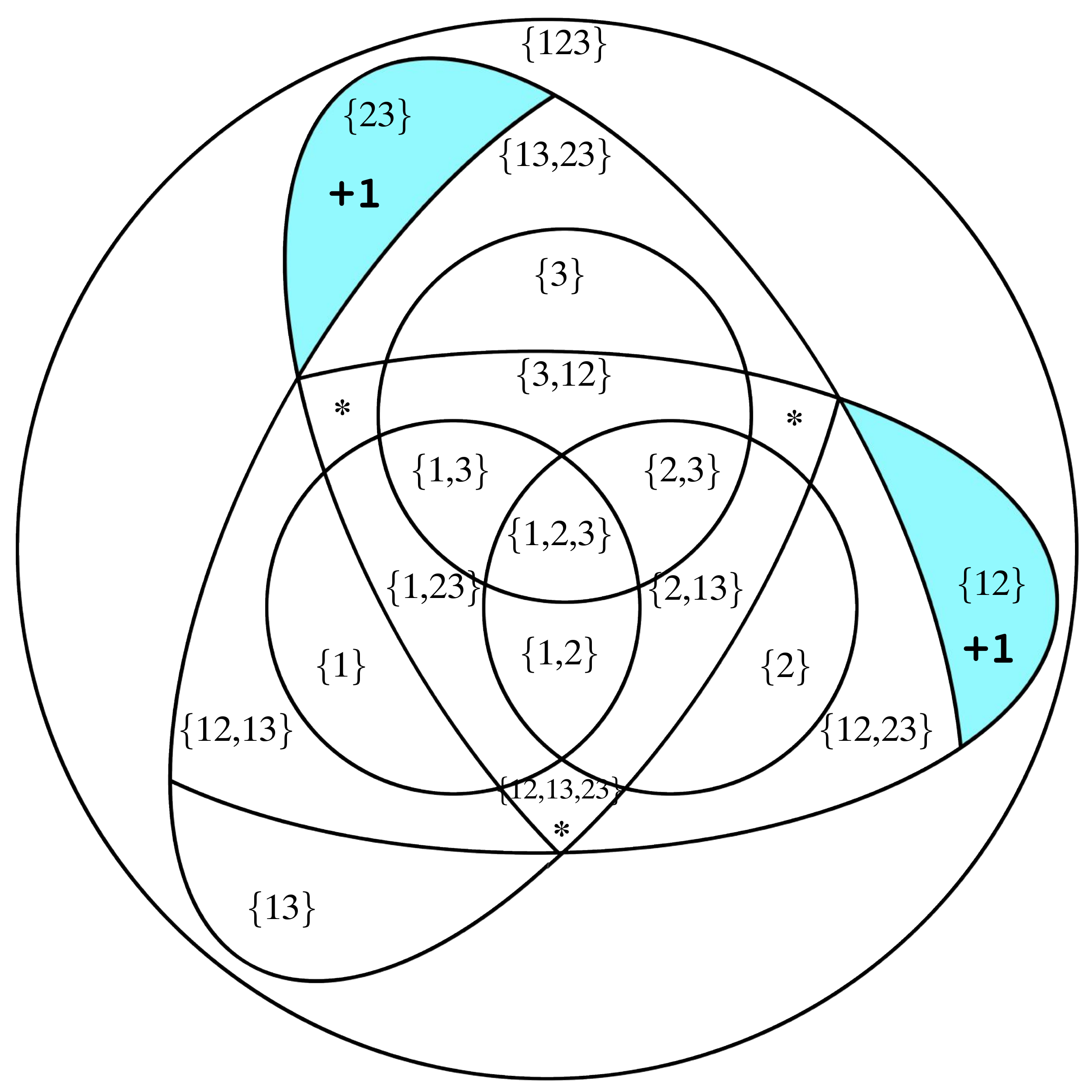} } \end{minipage}
	\caption{Example \textsc{DoubleXor}.  Target $Y$ has two bits of uncertainty.  The doublet $X_1 X_2$ specifies the ``left bit'' (\texttt{l}/\texttt{L}) and doublet $X_2 X_3$ specifies the ``right bit'' (\texttt{r}/\texttt{R}) for a total of $\info{X_1 X_2}{Y} + \info{X_2 X_3}{Y} = \ent{Y} = 2$ bits.  $X_1 X_2 X_3$'s specification of $Y$ is irreducible to disjoint parts yet fully reduces to the pair of parts $\{X_1 X_2, X_2 X_3\}$.}
	\label{fig:DoubleXor}
\end{figure}

In example \textsc{DoubleXor} (\figref{fig:DoubleXor}) the whole specifies two bits, $\info{X_1 X_2 X_3}{Y}~=~\ent{Y} = 2$ bits.  The doublet $X_1 X_2$ solely specifies the ``left-bit'', and the doublet $X_2 X_3$ solely specifies the ``right-bit''.  Applying each notion of irreducibility to \textsc{DoubleXor}:

\begin{enumerate}
\item[\textbf{\textsf{IbE}}] How much of $X_1 X_2 X_3$'s information about $Y$ can be reduced to the information conveyed by singleton elements?  The three singleton elements specify nothing about $Y$, $\info{X_i}{Y}~=~0$ bits $\forall i$.  This means the whole is utterly irreducible to its elements, making $\IbS\left( \setX : Y \right) = \info{X_1 X_2 X_3}{Y} - 0 = 2$ bits.

\item[\textbf{\textsf{IbDp}}]	How much of $X_1 X_2 X_3$'s information about $Y$ can be reduced to the information conveyed by disjoint parts?  Per subfigures \ref{fig:PID3_COPa}--\subref*{fig:PID3_COPc}, the three bipartitions of $X_1 X_2 X_3$ are: $\{X_1 X_2, X_3\}$, $\{X_1 X_3, X_2\}$, and $\{X_2 X_3, X_1\}$.  In the first bipartition, $\{X_1 X_2, X_3\}$, the doublet $X_1 X_2$ specifies the left-bit of $Y$ and the singleton $X_3$ specifies nothing for a total of one bit.  Similarly, in the second bipartition, $\{X_2 X_3, X_1\}$, $X_2 X_3$ specifies the right-bit of $Y$ and the singleton $X_1$ specifies nothing for a total of one bit.  Finally, in the bipartition $\{X_2 X_3, X_1\}$ both $X_2 X_3$ and $X_1$ specify nothing for a total of zero bits.  Taking the maximum over the three bipartitions, $\max[1,1,0]=1$, we discover disjoint parts specify at most one bit, leaving $\IbP\!\left( \setX : Y \right) =\info{X_1 X_2 X_3}{Y} - 1 = 2 - 1 = 1$ bit.

\item[\textbf{\textsf{Ib2p}}]	How much of $X_1 X_2 X_3$'s information about $Y$ can be reduced to the information conveyed by two parts?  Per subfigures \ref{fig:PID3_Ib2a}--\subref*{fig:PID3_Ib2c}, there are three pairs of Almosts, and one of them is $\{X_1 X_2, X_1 X_3\}$.  The Almost $X_1 X_2$ specifies the left-bit of $Y$, and the Almost $X_1 X_3$ specifies the right-bit of $Y$.  As there is a pair of parts that fully accounts for $X_1 X_2 X_3$'s specification of $Y$, $\IbTwo\!\left( \setX : Y \right) = 0$ bits.

	\item[\textbf{\textsf{IbAp}}] How much of $X_1 X_2 X_3$'s information about $Y$ can be reduced to the information conveyed by all possible parts?  From above we see that $\IbTwo$ is zero bits.  Per eq.~\eqref{eq:ordering}, $\IbD$ is stricter than $\IbC$, therefore $\IbD$ is also zero bits.
\end{enumerate}
	
\subsection{TripleXor: Irreducible to a pair of components, yet still reducible}

\begin{figure}[tbh]
	\centering
\begin{minipage}[l]{0.4\linewidth}
	\vspace{0.73in}	
	\subfloat[circuit diagram]{ \includegraphics[width=2.2in]{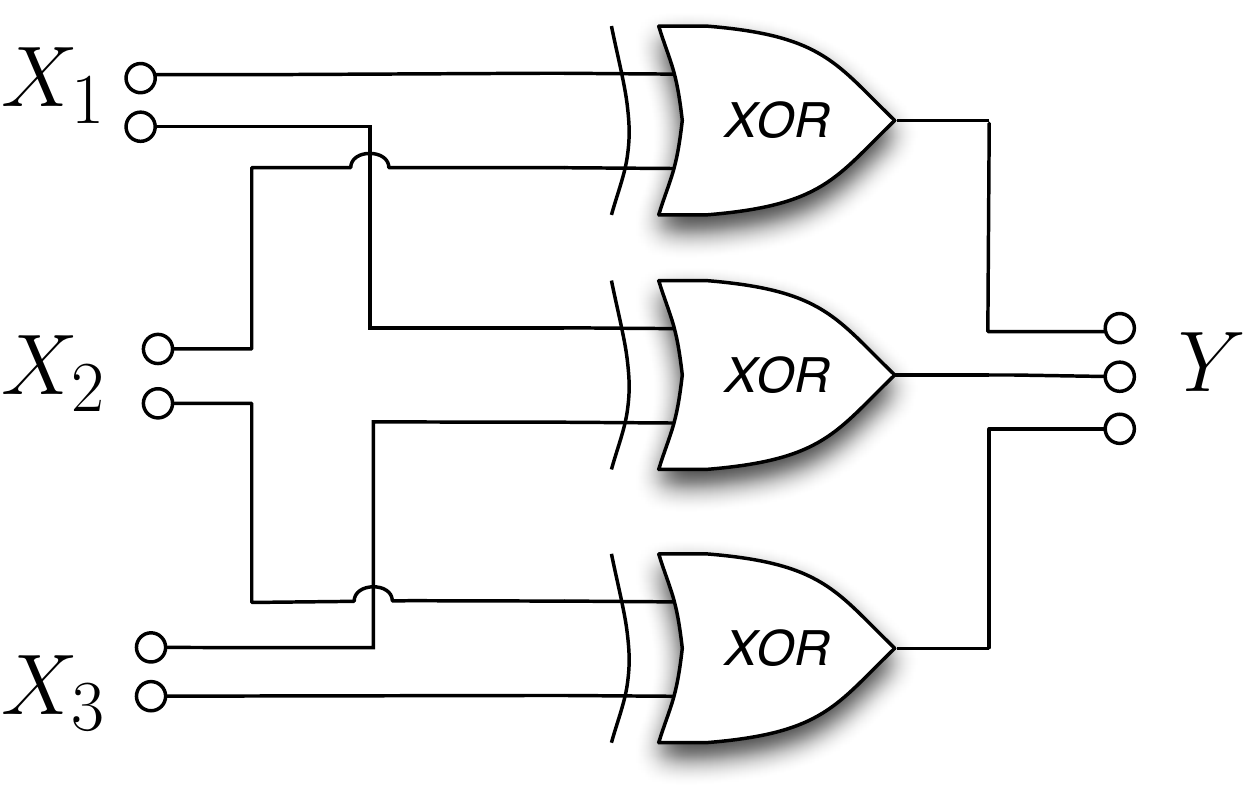} } \\
	\subfloat[$\Prob{x_1,x_2,x_3,y}$]{See Appendix B for the joint distribution. } \\
\end{minipage}	\begin{minipage}[r]{0.5\linewidth} \centering
	\subfloat[PI-diagram]{ \includegraphics[height=2.3in]{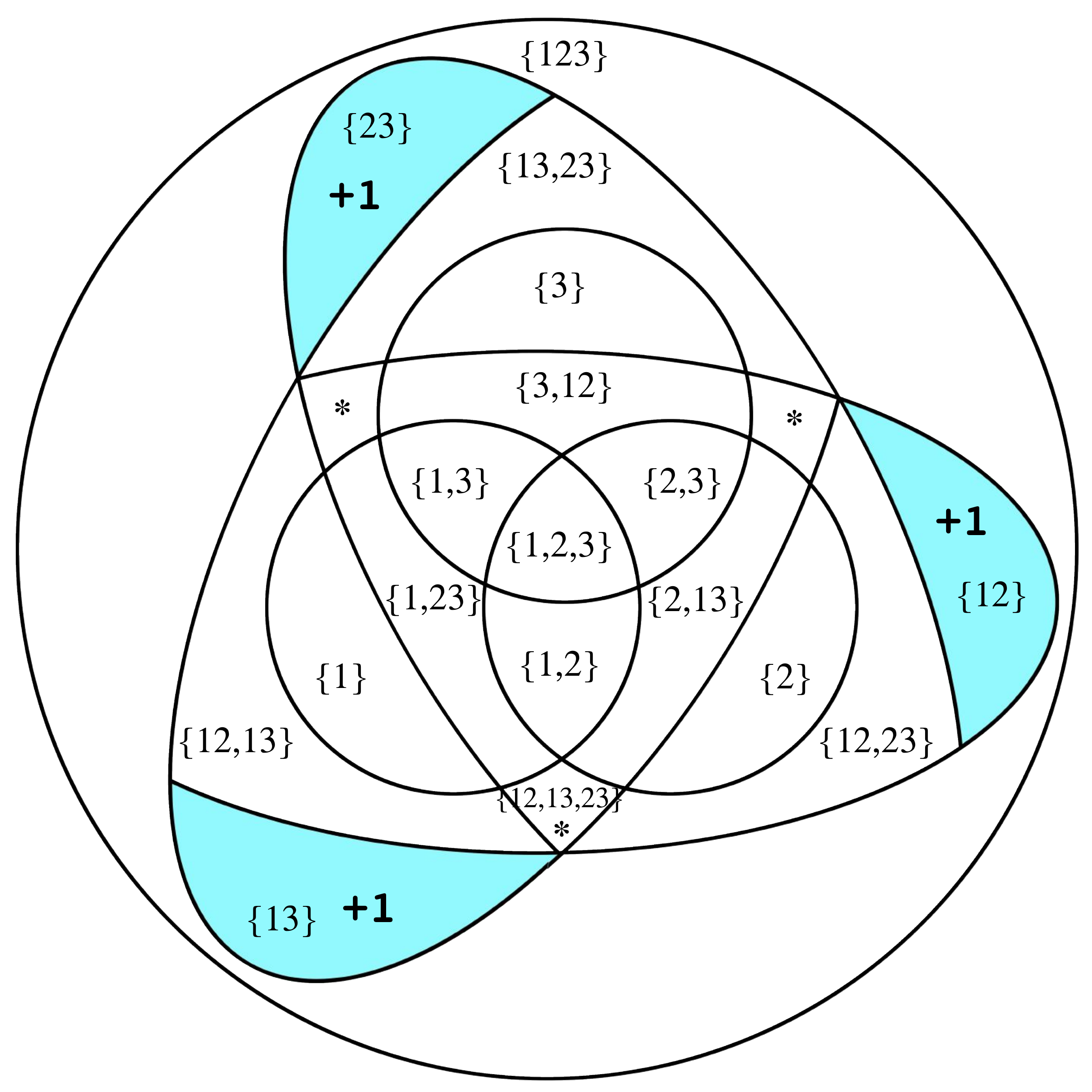} } \end{minipage}
	\caption{Example \textsc{TripleXor}.  Target $Y$ has three bits of uncertainty.  Each doublet part of $X_1 X_2 X_3$ specifies a distinct bit of $Y$, for a total of $\info{X_1 X_2}{Y} + \info{X_1 X_3}{Y} + \info{X_2 X_3}{Y} = \ent{Y} = 3$ bits.  The whole's specification of $Y$ is irreducible to any pair of Almosts yet fully reduces to all Almosts.}
	\label{fig:TripleXor}
\end{figure}

Example \textsc{TripleXor} (\figref{fig:TripleXor}) has trifold symmetry and the whole specifies three bits, \newline $\info{X_1 X_2 X_3}{Y} = \ent{Y} = 3$ bits.  Each bit is solely specified by one of three doublets: $X_1 X_2$, $X_1 X_3$, or $X_2 X_3$.  Applying each notion of irreducibility to \textsc{TripleXor}:

\begin{enumerate}
\item[\textbf{\textsf{IbE}}]	Working individually, the three elements specify absolutely nothing about $Y$, \newline $\info{X_1}{Y} = \info{X_2}{Y} = \info{X_3}{Y} = 0$ bits.  Thus the whole is utterly irreducible to elements, making $\IbS\left( \setX : Y \right) = \info{X_1 X_2 X_3}{Y} - 0 =3$ bits.

\item[\textbf{\textsf{IbDp}}]	The three bipartitions of $X_1 X_2 X_3$ are: $\{X_1 X_2, X_3\}$, $\{X_1 X_3, X_2\}$, and $\{X_2 X_3, X_1\}$.  In the first bipartition, $\{X_1 X_2, X_3\}$, the doublet $X_1 X_2$ specifies one bit of $Y$ and the singleton $X_3$ specifies nothing for a total of one bit.  By \textsc{TripleXor}'s trifold symmetry, we get the same value for bipartitions $\{X_1 X_2, X_3\}$ and $\{X_2 X_3, X_1\}$.  Taking the maximum over the three bipartitions, $\max[1,1,1]=1$, we discover a partition specifies at most one bit, leaving $\IbP\left( \setX : Y \right) = \info{X_1 X_2 X_3}{Y} - 1 = 2$ bits.

\item[\textbf{\textsf{Ib2p}}] There are three pairs of Almosts; they are: $\{X_1 X_2, X_2 X_3\}$, $\{X_1 X_2, X_1 X_3\}$, and $\{X_1 X_3, X_2 X_3\}$.  Each pair of Almosts specifies exactly two bits.  Taking the maximum over the pairs, $\max[2,2,2]=2$, we discover a pair of parts specifies at most two bits, leaving \newline $\IbC\left( \setX : Y \right) = \info{X_1 X_2 X_3}{Y} - 2 = 3 - 2 = 1$ bit.

\item[\textbf{\textsf{IbAp}}]	The $n$ Almosts of $X_1 X_2 X_3$ are $\{X_1, X_2, X_1 X_3, X_2 X_3\}$.  Each Almost specifies one bit of $Y$, for a total of three bits, making $\IbD\left( \setX : Y \right) =\info{X_1 X_2 X_3}{Y} - 3 = 0$ bits.
\end{enumerate}
	
\subsection{Parity: Complete irreducibility}
\begin{figure}[h!bt]
	\centering
	\begin{minipage}[c]{0.60\linewidth} \centering
	\subfloat[circuit diagram]{ \includegraphics[height=1.0in]{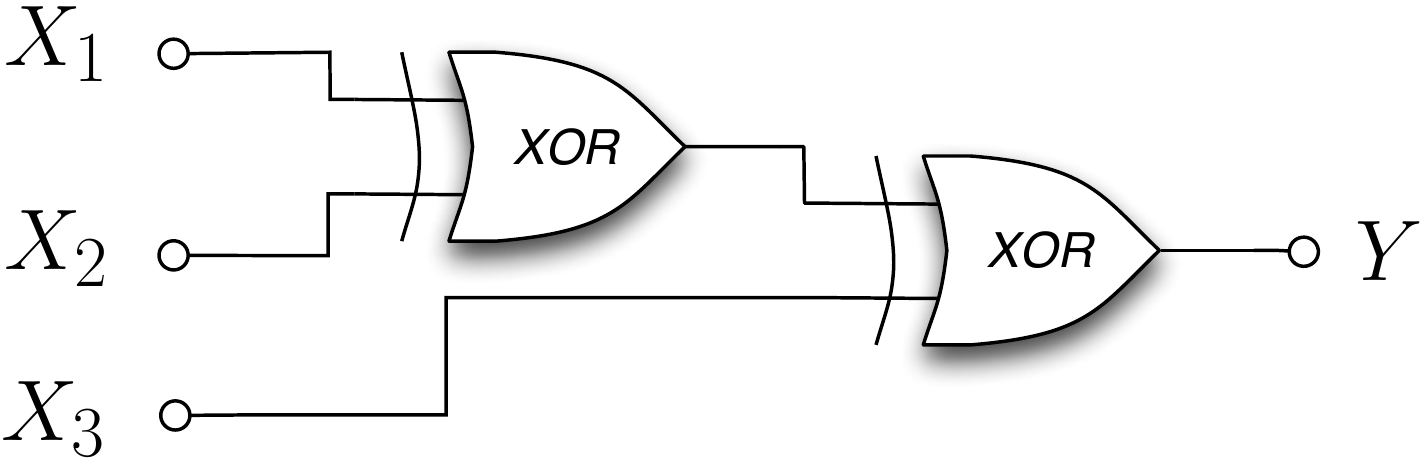} }
	\end{minipage}
	
	\begin{minipage}[c]{0.35\linewidth} \centering	
	\subfloat[$\Prob{x_1, x_2, x_3, y}$]{ \begin{tabular}{ c | c c } \cmidrule(r){1-2}
$\ \, X_1 \, X_2 \, X_3$  &$Y$ \\
\cmidrule(r){1-2} 
\bin{0 0 0} & \bin{0} & \quad \nicefrac{1}{8}\\
\bin{0 0 1} & \bin{1} & \quad \nicefrac{1}{8}\\
\bin{0 1 0} & \bin{1} & \quad \nicefrac{1}{8}\\
\bin{0 1 1} & \bin{0} & \quad \nicefrac{1}{8}\\
\bin{1 0 0} & \bin{1} & \quad \nicefrac{1}{8}\\
\bin{1 0 1} & \bin{0} & \quad \nicefrac{1}{8}\\
\bin{1 1 0} & \bin{0} & \quad \nicefrac{1}{8}\\
\bin{1 1 1} & \bin{1} & \quad \nicefrac{1}{8}\\
\cmidrule(r){1-2} 
\end{tabular} }
\end{minipage} \begin{minipage}[c]{0.55\linewidth} \centering
	\subfloat[PI-diagram]{ \includegraphics[height=2.3in]{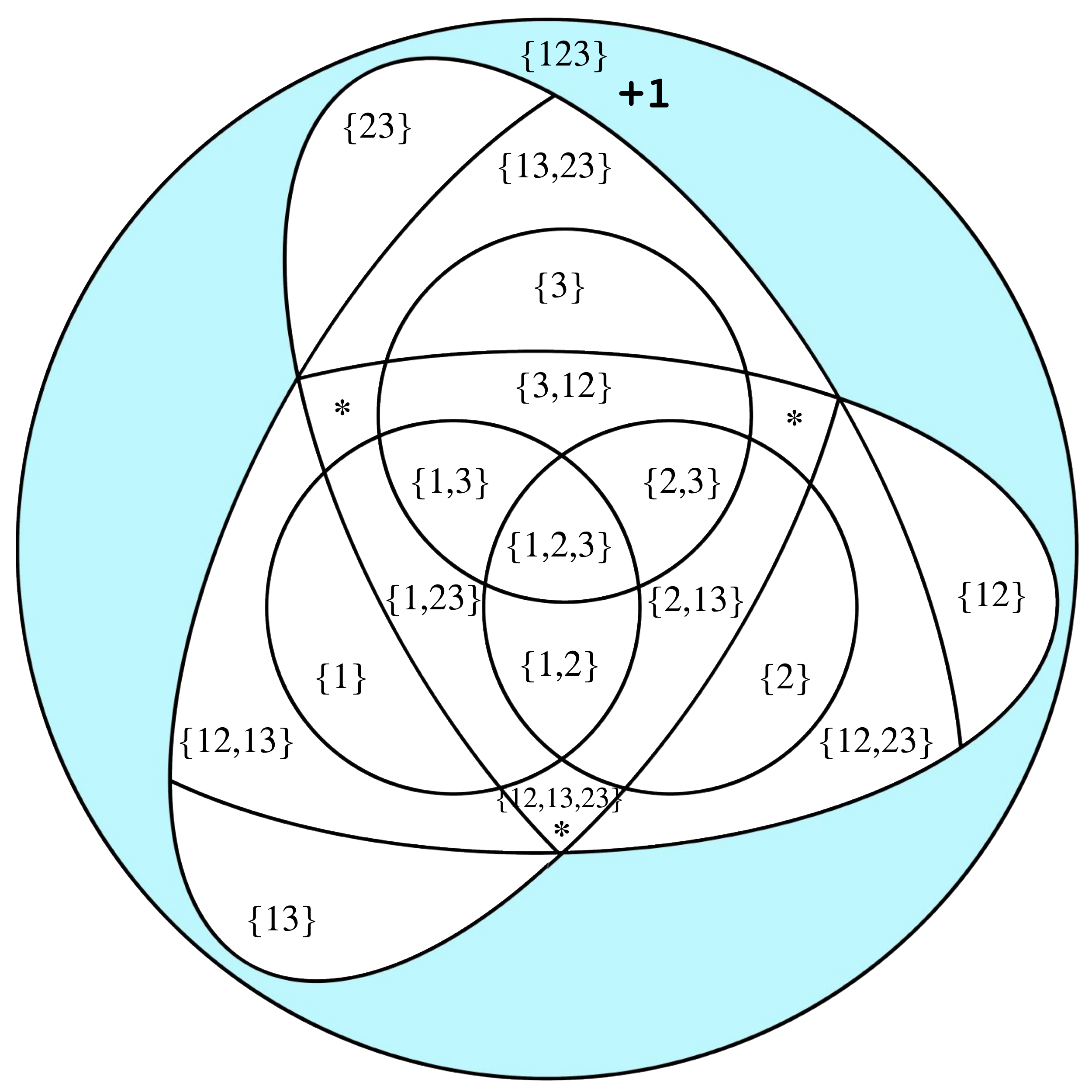} }
	\end{minipage}
	\caption{Example \textsc{Parity}.  Target $Y$ has one bit of uncertainty, and only the whole specifies $Y$, $\info{X_1 X_2 X_3}{Y}~=~\ent{Y}~=~1$ bit.  $X_1 X_2 X_3$'s specification of $Y$ is utterly irreducible to any collection of $X_1 X_2 X_3$'s parts, and $\IbD( \{X_1, X_2, X_3\} : Y)~=~1$ bit.}
	\label{fig:Parity}
\end{figure}

In example \textsc{Parity} (\figref{fig:Parity}), the whole specifies one bit of uncertainty, $\info{X_1 X_2 X_3}{Y}~=~\ent{Y}~=~1$ bit.  No singleton or doublet specifies anything about $Y$, $\info{X_i}{Y}~=~\info{X_i X_j}{Y}~=~0$ bits $\forall i,j$.  Applying each notion of irreducibility to \textsc{Parity}:

\begin{enumerate}
\item[\textbf{\textsf{IbE}}] The whole specifies one bit, yet the elements $\{X_1, X_2, X_3\}$ specify nothing about $Y$.  Thus the whole is utterly irreducible to elements making, $\IbS\left( \setX : Y \right) = \info{X_1 X_2 X_3}{Y} - 0 =1$ bit.

\item[\textbf{\textsf{IbDp}}] The three bipartitions of $\setX$ are: $\{X_1 X_2, X_3\}$, $\{X_1 X_3, X_2\}$, and $\{X_2 X_3, X_1\}$.  By the above each doublet and singleton specifies nothing about $Y$, and thus each partition specifies nothing about $Y$.  Taking the maximum over the bipartitions yields $\max[0,0,0]=0$, making \newline $\IbP( \setX : Y ) ~=~1~-~0 =1$ bit.

\item[\textbf{\textsf{Ib2p}}] The pairs of $\setX$'s Almosts are: $\{X_1 X_2, X_1 X_3\}$, $\{X_1 X_2, X_2 X_3\}$, and $\{X_1 X_3, X_2 X_3\}$.  As before, each doublet specifies nothing about $Y$, and a pair of nothings is still nothing.  Taking the maximum yields $\max[0,0,0]=0$, making $\IbC( \setX : Y )~=~1~-~0 =1$ bit.

\item[\textbf{\textsf{IbAp}}] The three Almosts of $\setX$ are: $\{X_1 X_2, X_1 X_3, X_2 X_3\}$.  Each Almost specifies nothing, and a triplet of nothings is still nothing, making $\IbD( \setX : Y)~=~1~-~0~=~1$ bit.
\end{enumerate}

Table \ref{fig:thetable} summarizes the results of our four irreducibility measures applied to our examples.

\begin{table}[hbt]
\centering
	\begin{tabular}{ l c c c c c } \toprule
 \addlinespace
		Example & $\info{\Xn}{Y}$ & $\IbS$ & $\IbP$ & $\IbTwo$ & $\IbD$ \\
	\midrule
\textsc{Xor} (Fig. \ref{fig:XOR}) & 1 & 1 & 1 & 1 & 1 \vspace{3pt} \\
\textsc{XorUnique} (Fig. \ref{fig:XorUnique}) & 2 & 1 & 0 & 0 & 0 \vspace{3pt} \\
\textsc{DoubleXor} (Fig. \ref{fig:DoubleXor}) & 2 & 2 & 1 & 0 & 0 \vspace{3pt} \\
\textsc{TripleXor} (Fig. \ref{fig:TripleXor}) & 3 & 3 & 2 & 1 & 0 \vspace{3pt} \\
\textsc{Parity}    (Fig. \ref{fig:Parity}) & 1 & 1 & 1 & 1 & 1 \\
\bottomrule
\end{tabular}
\caption{Irreducibility values for our exemplary binary circuits.}
\label{fig:thetable}
\end{table}

\section{Conclusion}
Within the Partial Information Decomposition framework\cite{plw-10}, synergy the simplest case of the broader notion of irreducibility.  PI-diagrams, a generalization of Venn diagrams, are immensely helpful in improving one's intuition for synergy and irreducibility.

We define the irreducibility of the mutual information a set of $n$ random variables $\setX = \{X_1, \ldots, X_n\}$ convey about a target $Y$ as the information the whole conveys about $Y$, $\info{\Xn}{Y}$, minus the maximum union-information conveyed by the ``parts'' of $\setX$.  The four common notions of $\setX$'s parts are: (1) the set of the $n$ atomic elements; (2) all partitions of disjoint parts; (3) all pairs of parts; and (4) the set of all $2^{n} - 2$ possible parts.  All four definitions of parts are equivalent when the whole consists of two atomic elements $(n=2)$, but they diverge for $n > 2$.  We anticipate this work will become more useful once the complexity community has converged on a palatable $\Icap$ measure.

\bibliography{quant_synergy}

\appendix


\section{Proofs}

{\lem{We prove that Information beyond the Bipartition, $\IbPTwo( \setX : Y )$, equals Information beyond the Disjoint Parts, $\IbP( \setX : Y )$ by showing,
\[
    \IbP( \setX : Y ) \leq \IbPTwo( \setX : Y ) \leq \IbP( \setX : Y ) \; .
\]
\label{proof:IbP} }}

\begin{proof}
We first show that $\IbP( \setX : Y) \leq \IbPTwo( \setX : Y )$.  By their definitions:
\begin{eqnarray}
    \IbP( \setX : Y ) &\equiv& \info{Y}{\Xn} - \max_{\mathbf{P}} \opI_{\cup}\left( Y : \mathbf{P} \right) \\
    \IbPTwo( \setX : Y ) &\equiv& \info{Y}{\Xn} - \max_{S \subset \setX} \opI_{\cup}\left( Y : \{S, \setX \setminus S \} \right) \\
    &=& \info{Y}{\Xn} - \max_{ \substack{\mathbf{P} \\ |\mathbf{P}| = 2} } \opI_{\cup}\left( Y : \mathbf{P} \right) \; ,
\end{eqnarray}
where $\P$ enumerates over all disjoint parts of $\setX$.

By removing the restriction that $|\mathbf{P}|=2$ from the minimized union-information in $\IbPTwo$ we arrive at $\IbP$.    As removing a restriction can only decrease the minimum, therefore \newline $\IbP( \setX~:~Y)~\leq~\IbPTwo(\setX~:~Y)$.
\end{proof}

We next show that $\IbPTwo( \setX : Y ) \leq \IbP\left( \setX : Y \right)$.  Meaning we must show that, 

\begin{equation}
    \label{eq:rightbeforeprovingIbB}
    \info{\Xn}{Y} - \max_{ \substack{\mathbf{P} \\ |\mathbf{P}| = 2} } \Icupe{\setP}{Y}  \leq \info{\Xn}{Y} - \max_{\mathbf{P}} \Icupe{\setP}{Y} \; ,
\end{equation}
where $\P$ enumerates over all disjoint parts of $\setX$.

\begin{proof}
By subtracting $\info{\Xn}{Y}$ from each side and multiplying each side by $-1$ we have,
\begin{equation}
    \label{eq:IbPunion1}
    \max_{ \substack{\mathbf{P} \\ |\mathbf{P}| = 2} } \opI_{\cup}\left( \mathbf{P} : Y \right) \geq \max_{\mathbf{P}}  \; \opI_{\cup}\left( \mathbf{P} : Y \right) \; .
\end{equation}

Without loss of generality, we take any individual subset/part $S$ in $X$.  Then we have a bipartition $\mathbf{B}$ of parts $\{S, \setX \setminus S\}$.  We then further partition the part $\setX \setminus S$ into $k$ disjoint subcomponents denoted $\{T_1, \ldots, T_k\}$ where $2 \leq k \leq n - \left| S \right|$ creating an arbitrary partition $\mathbf{P}=\{S, T_1, \ldots, T_k\}$.  We now need to show that,

\begin{equation}
    \Icupe{S, \setX \setminus S}{Y} \geq \Icupe{S, T_1, \ldots, T_k}{Y} \; .
\end{equation}

By the monotonicity axiom \textbf{(M)}, we can append each subcomponent $T_1, \ldots, T_k$ to $\mathbf{B}$ without changing the union-information because every subcomponent $T_i$ is a subset of the element $\setX \setminus S$.  Then using the symmetry axiom $\mathbf{(S_0)}$, we re-order the parts so that $S, T_1, \ldots, T_k$ come first.  This yields,

\begin{equation}
    \Icupe{S, T_1, \ldots, T_k, \setX \setminus S}{Y} \geq \Icupe{S, T_1, \ldots, T_k}{Y} \; .
\end{equation}

Applying the monotonicity axiom \textbf{(M)} again, we know that adding the entry $\setX \setminus S$ can only increase the union information.  Therefore we prove eq.~\eqref{eq:IbPunion1}, which proves eq.~\eqref{eq:rightbeforeprovingIbB}.
\end{proof}

Finally, by the squeeze theorem we prove Lemma \ref{proof:IbP}

{\lem{Proof that pairs of Almosts cover \IbC.  We prove that the maximum union-information over all possible pairs of parts $\{P_1, P_2\}$, equates to the maximum union-information over all pairs of Almosts $\{A_i, A_j\} \ i \not= j$.  Mathematically,

\begin{equation}
    \max_{\substack{P_1, P_2 \\ P_1, P_2 \subset \setX}} \Icupe{P_1, P_2}{Y} = \max_{ \substack{i,j \in \{1, \ldots, n\} \\ i \not= j} } \Icupe{A_i, A_j}{Y} \; . 
\end{equation} \label{proof:IbC} } }

\begin{proof}
By the right-monotonicity lemma \textbf{(RM)}, the union-information can only increase when increasing the size of the parts $P_1$ and $P_2$.  We can therefore ignore all parts $P_1, P_2$ of size less than $n-1$,
\begin{eqnarray}
\max_{\substack{P_1, P_2 \\ P_1, P_2 \subset \setX}} \Icupe{P_1, P_2}{Y}&=& \max_{\substack{P_1, P_2 \\ P_1, P_2 \in \mathcal{P}(\setX) \\ |P_1| = |P_2| = n-1}} \Icupe{P_1, P_2}{Y} \\
    &=& \max_{ i,j \in \{1,\ldots, n\} } \Icupe{ A_i, A_j }{Y} \; .
\end{eqnarray}

Then by the idempotency axiom \textbf{(I)} and then the monotonicity axiom \textbf{(M)}, having $i \not= j$ can only increase the union information.  Therefore,
\begin{equation}
\label{eq:bipartequiv}
\max_{ i,j \in \{1,\ldots, n\} } \Icupe{A_i, A_j}{Y} = \max_{\substack{i,j \in \{1, \ldots, n\} \\ i \not= j}} \Icupe{A_i, A_j}{Y} \; .
\end{equation}

With eq.~\eqref{eq:bipartequiv} in hand, we easily show that the Information beyond all pairs of Subsets, $\IbC$, equates to the information beyond all pairs of Almosts,
\begin{eqnarray}
    \IbC\left( \setX : Y \right) &\equiv& \info{\Xn}{Y} -  \max_{\substack{P_1, P_2 \\ P_1, P_2 \in \mathcal{P}(\setX)}} \Icupe{P_1, P_2}{Y} \\
    &=& \info{\Xn}{Y} - \max_{\substack{i,j \in \{1, \ldots, n\} \\ i \not= j}} \Icupe{A_i, A_j}{Y} \; .
\end{eqnarray} 
\end{proof}

{\lem{Proof that Almosts cover \IbD.  We wish to show that the union-information over all distinct parts of $n$ elements, $\mathcal{P}(\setX)$, is equivalent to the union information over the $n$ Almosts.  Mathematically,
\begin{equation}
    \Icupe{\mathcal{P}(\setX)}{Y} = \Icupe{A_1, \ldots, A_n}{Y} \; .
\end{equation} \label{proof:IbD} } }

\begin{proof}
Every element in the set of parts $\mathcal{P}(\setX)$ that isn't an Almost is a subset of an Almost.  Therefore by the monotonicity axiom \textbf{(M)} we can remove this entry.  Repeating this process we remove all entries except the $n$ Almosts.  Therefore, $\opI_{\cup}\!\left( \mathcal{P}(\setX) : Y \right)~=~\Icupe{A_1, \ldots, A_n}{Y}$.
\end{proof}

\clearpage
\section{Joint distributions for DoubleXor and TripleXor}
\label{app:jointdistributions}

\begin{figure}[h!bt]
	\centering
\begin{tabular}{ c | c c} \cmidrule(r){1-2}
	$X_1$ $X_2$ $X_3$ &$Y$ \\
	\cmidrule(r){1-2} 
	\bin{0 00 0} & \bin{lr} & \quad \nicefrac{1}{16}\\
	\bin{0 01 0} & \bin{lR} & \quad \nicefrac{1}{16}\\
	\bin{0 10 0} & \bin{Lr} & \quad \nicefrac{1}{16}\\
	\bin{0 11 0} & \bin{LR} & \quad \nicefrac{1}{16}\\
\addlinespace
	\bin{0 00 1} & \bin{lR} & \quad \nicefrac{1}{16}\\
	\bin{0 01 1} & \bin{lr} & \quad \nicefrac{1}{16}\\
	\bin{0 10 1} & \bin{LR} & \quad \nicefrac{1}{16}\\
	\bin{0 11 1} & \bin{Lr} & \quad \nicefrac{1}{16}\\
\addlinespace
	\bin{1 00 0} & \bin{Lr} & \quad \nicefrac{1}{16}\\
	\bin{1 01 0} & \bin{LR} & \quad \nicefrac{1}{16}\\
	\bin{1 10 0} & \bin{lr} & \quad \nicefrac{1}{16}\\
	\bin{1 11 0} & \bin{lR} & \quad \nicefrac{1}{16}\\
\addlinespace
	\bin{1 00 1} & \bin{LR} & \quad \nicefrac{1}{16}\\
	\bin{1 01 1} & \bin{Lr} & \quad \nicefrac{1}{16}\\
	\bin{1 10 1} & \bin{lR} & \quad \nicefrac{1}{16}\\
	\bin{1 11 1} & \bin{lr} & \quad \nicefrac{1}{16}\\
	\cmidrule(r){1-2} 
\end{tabular}
	\caption{Joint distribution $\Prob{x_1, x_2, x_3, y}$ for example \textsc{DoubleXor}.}
	\label{fig:DoubleXorJointDist}
\end{figure}

\begin{figure}[h!bt]
	\centering
\begin{tabular}{ c | c c} \cmidrule(r){1-2}
	$X_1$ $X_2$ $X_3$ &$Y$ \\
\cmidrule(r){1-2} 
	\bin{00 00 00} & \bin{000} & \quad \nicefrac{1}{64}\\
	\bin{00 00 01} & \bin{001} & \quad \nicefrac{1}{64}\\
	\bin{00 00 10} & \bin{010} & \quad \nicefrac{1}{64}\\
	\bin{00 00 11} & \bin{011} & \quad \nicefrac{1}{64}\\
\addlinespace
	\bin{00 01 00} & \bin{001} & \quad \nicefrac{1}{64}\\
	\bin{00 01 01} & \bin{000} & \quad \nicefrac{1}{64}\\
	\bin{00 01 10} & \bin{011} & \quad \nicefrac{1}{64}\\
	\bin{00 01 11} & \bin{010} & \quad \nicefrac{1}{64}\\
\addlinespace
	\bin{00 10 00} & \bin{100} & \quad \nicefrac{1}{64}\\
	\bin{00 10 01} & \bin{101} & \quad \nicefrac{1}{64}\\
	\bin{00 10 10} & \bin{110} & \quad \nicefrac{1}{64}\\
	\bin{00 10 11} & \bin{111} & \quad \nicefrac{1}{64}\\
\addlinespace
	\bin{00 11 00} & \bin{101} & \quad \nicefrac{1}{64}\\
	\bin{00 11 01} & \bin{100} & \quad \nicefrac{1}{64}\\
	\bin{00 11 10} & \bin{111} & \quad \nicefrac{1}{64}\\
	\bin{00 11 11} & \bin{110} & \quad \nicefrac{1}{64}\\
\addlinespace
	\bin{01 00 00} & \bin{000} & \quad \nicefrac{1}{64}\\
	\bin{01 00 01} & \bin{001} & \quad \nicefrac{1}{64}\\
	\bin{01 00 10} & \bin{010} & \quad \nicefrac{1}{64}\\
	\bin{01 00 11} & \bin{011} & \quad \nicefrac{1}{64}\\
\addlinespace
	\bin{01 01 00} & \bin{001} & \quad \nicefrac{1}{64}\\
	\bin{01 01 01} & \bin{000} & \quad \nicefrac{1}{64}\\
	\bin{01 01 10} & \bin{011} & \quad \nicefrac{1}{64}\\
	\bin{01 01 11} & \bin{010} & \quad \nicefrac{1}{64}\\
\addlinespace
	\bin{01 10 00} & \bin{100} & \quad \nicefrac{1}{64}\\
	\bin{01 10 01} & \bin{101} & \quad \nicefrac{1}{64}\\
	\bin{01 10 10} & \bin{110} & \quad \nicefrac{1}{64}\\
	\bin{01 10 11} & \bin{111} & \quad \nicefrac{1}{64}\\
\addlinespace
	\bin{01 11 00} & \bin{101} & \quad \nicefrac{1}{64}\\
	\bin{01 11 01} & \bin{100} & \quad \nicefrac{1}{64}\\
	\bin{01 11 10} & \bin{111} & \quad \nicefrac{1}{64}\\
	\bin{01 11 11} & \bin{110} & \quad \nicefrac{1}{64}\\
\cmidrule(r){1-2} 
\end{tabular} \qquad \qquad \qquad \qquad \begin{tabular}{ c | c c} \cmidrule(r){1-2}
	$X_1$ $X_2$ $X_3$ &$Y$ \\
\cmidrule(r){1-2} 
	\bin{10 00 00} & \bin{110} & \quad \nicefrac{1}{64}\\
	\bin{10 00 01} & \bin{111} & \quad \nicefrac{1}{64}\\
	\bin{10 00 10} & \bin{100} & \quad \nicefrac{1}{64}\\
	\bin{10 00 11} & \bin{101} & \quad \nicefrac{1}{64}\\
\addlinespace
	\bin{10 01 00} & \bin{111} & \quad \nicefrac{1}{64}\\
	\bin{10 01 01} & \bin{110} & \quad \nicefrac{1}{64}\\
	\bin{10 01 10} & \bin{101} & \quad \nicefrac{1}{64}\\
	\bin{10 01 11} & \bin{100} & \quad \nicefrac{1}{64}\\
\addlinespace
	\bin{10 10 00} & \bin{010} & \quad \nicefrac{1}{64}\\
	\bin{10 10 01} & \bin{011} & \quad \nicefrac{1}{64}\\
	\bin{10 10 10} & \bin{000} & \quad \nicefrac{1}{64}\\
	\bin{10 10 11} & \bin{001} & \quad \nicefrac{1}{64}\\
\addlinespace
	\bin{10 11 00} & \bin{011} & \quad \nicefrac{1}{64}\\
	\bin{10 11 01} & \bin{010} & \quad \nicefrac{1}{64}\\
	\bin{10 11 10} & \bin{001} & \quad \nicefrac{1}{64}\\
	\bin{10 11 11} & \bin{000} & \quad \nicefrac{1}{64}\\
\addlinespace
	\bin{11 00 00} & \bin{110} & \quad \nicefrac{1}{64}\\
	\bin{11 00 01} & \bin{111} & \quad \nicefrac{1}{64}\\
	\bin{11 00 10} & \bin{100} & \quad \nicefrac{1}{64}\\
	\bin{11 00 11} & \bin{101} & \quad \nicefrac{1}{64}\\
\addlinespace
	\bin{11 01 00} & \bin{011} & \quad \nicefrac{1}{64}\\
	\bin{11 01 01} & \bin{010} & \quad \nicefrac{1}{64}\\
	\bin{11 01 10} & \bin{001} & \quad \nicefrac{1}{64}\\
	\bin{11 01 11} & \bin{000} & \quad \nicefrac{1}{64}\\
\addlinespace
	\bin{11 10 00} & \bin{010} & \quad \nicefrac{1}{64}\\
	\bin{11 10 01} & \bin{011} & \quad \nicefrac{1}{64}\\
	\bin{11 10 10} & \bin{000} & \quad \nicefrac{1}{64}\\
	\bin{11 10 11} & \bin{001} & \quad \nicefrac{1}{64}\\
\addlinespace
	\bin{11 11 00} & \bin{011} & \quad \nicefrac{1}{64}\\
	\bin{11 11 01} & \bin{010} & \quad \nicefrac{1}{64}\\
	\bin{11 11 10} & \bin{001} & \quad \nicefrac{1}{64}\\
	\bin{11 11 11} & \bin{000} & \quad \nicefrac{1}{64}\\
\cmidrule(r){1-2} 
\end{tabular}
	\caption{Joint distribution $\Prob{x_1, x_2, x_3, y}$ for example \textsc{TripleXor}. }
	\label{fig:TripleXorJointDist}
\end{figure}

\end{document}